\documentclass[prd,twocolumn,nofootinbib,showpacs,superscriptaddress]{revtex4}

\usepackage{amsfonts}
\usepackage{amsmath}
\usepackage{amssymb}
\usepackage{bm}
\usepackage{dcolumn}
\usepackage{epsfig}
\usepackage{graphicx}
\usepackage{graphics}
\usepackage[latin1]{inputenc}
\usepackage{latexsym}
\usepackage{rotating}
\usepackage{hyperref}
\usepackage[usenames]{color}
\usepackage{float}
\usepackage{xspace} 
\usepackage{mathrsfs}
\usepackage{subfigure}
\usepackage{enumitem}
\usepackage{ulem}
\definecolor {darkgreen}{rgb}{0.2,0.7,0.2}

\newcommand\be{\begin{equation}}
\newcommand\ba{\begin{eqnarray}}
\newcommand\ee{\end{equation}}
\newcommand\ea{\end{eqnarray}}

\newcommand{\nn}{\nonumber}

\newcommand{\ST}{{\mbox{\tiny ST}}}

\newcommand{\BD}{{\mbox{\tiny BD}}}

\newcommand{\fmin}{{\mbox{\tiny min}}}

\newcommand{\se}{{\mbox{\tiny se}}}
\newcommand{\sn}{{\mbox{\tiny sn}}}
\newcommand{\br}{{\mbox{\tiny b}}}
\newcommand{\Lo}{{\mbox{\tiny L}}}

\newcommand{\inj}{{\mbox{\tiny inj}}}

\newcommand{\ppE}{{\mbox{\tiny ppE}}}

\newcommand{\GR}{{\mbox{\tiny GR}}}

\newcommand{\ISCO}{{\mbox{\tiny ISCO}}}

\begin{document}
\title{Projected Constraints on Scalarization \\ with Gravitational Waves from Neutron Star Binaries}

\author{Laura Sampson}
\affiliation{Department of Physics, Montana State University, Bozeman, MT 59717, USA.}

\author{Nicol\'as Yunes}
\affiliation{Department of Physics, Montana State University, Bozeman, MT 59717, USA.}

\author{Neil Cornish}
\affiliation{Department of Physics, Montana State University, Bozeman, MT 59717, USA.}

\author{Marcelo Ponce}
\affiliation{Department of Physics, University of Guelph, Guelph, Ontario N1G 2W1, Canada.}

\author{Enrico Barausse}
\affiliation{CNRS, UMR 7095, Institut d'Astrophysique de Paris, 98bis Bd Arago, 75014 Paris, France}
\affiliation{Sorbonne Universit\'es, UPMC Univ Paris 06, UMR 7095, 98bis Bd Arago, 75014 Paris, France}

\author{Antoine Klein}
\affiliation{Department of Physics and Astronomy, The University of Mississippi, University, MS 38677, USA}

\author{Carlos Palenzuela}
\affiliation{Canadian Institute for Theoretical Astrophysics, Toronto, Ontario M5S 3H8, Canada.}

\author{Luis Lehner}
\affiliation{Perimeter Institute for Theoretical Physics, Waterloo, Ontario N2L 2Y5, Canada.}
\affiliation{CIFAR, Cosmology \& Gravity Program, Toronto, ON M5G 1Z8, Canada Canada.}

\date{\today}

\begin{abstract} 

Certain scalar-tensor theories have the property of endowing stars with scalar hair, sourced either by the star's own compactness (spontaneous scalarization) or, for binary systems, by the companion's scalar hair (induced scalarization) or by the orbital binding energy (dynamical scalarization). Scalarized stars in binaries present different conservative dynamics than in General Relativity, and can also
excite a scalar mode in the metric perturbation that carries away dipolar radiation. 
As a result, the binary orbit shrinks faster than predicted in General Relativity, modifying the rate of decay of the orbital period. 
In spite of this, scalar-tensor theories can pass existing binary pulsar tests, because observed pulsars may not be compact enough or sufficiently orbitally bound to activate scalarization. 
Gravitational waves emitted during the last stages of compact binary inspirals are thus ideal probes of scalarization effects.
For the standard projected sensitivity of advanced LIGO, we here show that, if the neutron star equation of state is such that the stars can be sufficiently compact that they enter the detector's sensitivity band already scalarized, then gravitational waves could place constraints at least comparable to binary pulsars.
If the stars dynamically scalarize while inspiraling in band, then constraints are still possible provided the equation of state leads to scalarization that occurs sufficiently early in the inspiral, roughly below an orbital frequency of 50Hz.
In performing these studies, we re-derive an easy-to-calculate data analysis measure, an integrated phase difference between a General Relativistic and a modified signal, and connect it directly to the Bayes factor, showing that it can be used to determine whether a modified gravity effect is detectable. 
Finally, we find that custom-made templates are equally effective as model-independent, parameterized post-Einsteinian waveforms at detecting such modified gravity effects at realistic signal-to-noise ratios.
\end{abstract}

\maketitle
\allowdisplaybreaks[4]

\section{Introduction}

When gravitational waves (GWs) are detected by second-generation detectors (such as advanced LIGO (aLIGO)~\cite{ligo,Abramovici:1992ah,Abbott:2007kv}, advanced Virgo (aVirgo)~\cite{virgo,Giazotto:1988gw}, and KAGRA~\cite{KAGRA2}) some time in the next few years, one of the most exciting prospects is using these signals to test General Relativity (GR) in the very strong-field, \emph{dynamical and non-linear} regime~\cite{Yunes:2013dva}. There has been much work done on constraining departures from GR dynamics with Solar System and binary pulsar observations, and quite strong bounds have been placed on deviations from Einstein's theory in certain regimes. In particular, the strength of dipole radiation from some types of scalar-tensor (ST) theories is already tightly constrained by observations of the rate of decay of the orbital period of binary pulsars~\cite{Psaltis:2005ai, stairs, damour-taylor, kramer-wex, Freire:2012mg}. 
 
There remains, however, a class of ST theories that can escape these constraints. For theories in this class, initially proposed by Ref.~\cite{Damour:1992we,Damour:1993hw}, the scalar charge of a compact object is dependent upon the gravitational binding energy (or compactness) of the object itself (\emph{spontaneous scalarization})~\cite{Damour:1992we,Damour:1993hw}, and, if the object is in a binary, upon the orbital binding energy of the binary system (\emph{dynamical scalarization}), a phenomenon discovered in Ref.~\cite{Barausse:2012da}. Additionally, once a star acquires a scalar charge, it can scalarize its binary companion (\emph{induced scalarization})~\cite{Damour:1992we,Damour:1993hw,Damour:1996ke,Barausse:2012da,Palenzuela:2013hsa}.

These three types of scalarization can be understood in analogy with magnetization~\cite{Damour:1996ke}. Induced scalarization~\cite{Damour:1992we,Damour:1993hw,Damour:1996ke,Barausse:2012da,Palenzuela:2013hsa} is similar to what occurs in paramagnetism, where the individual magnetic moments of a large collection of atoms align themselves in the presence of a strong, external magnetic field. This alignment induces an overall magnetization of the collection of atoms. In the case of a neutron star (NS) in the presence of an external scalar field, e.g.~one supported by its binary companion, the star can develop a scalar field of its own.

Similarly, spontaneous~\cite{Damour:1992we,Damour:1993hw,Damour:1996ke} and dynamical~\cite{Barausse:2012da,Palenzuela:2013hsa} scalarization can both be understood in the context of spontaneous magnetization. In this phenomenon, a collection of unaligned magnetic moments will spontaneously align themselves in some direction as the temperature is lowered past a critical point, even in the absence of an external field. This occurs because, when the temperature is low enough, a new energy minimum appears -- a broken symmetry state that is associated with a non-zero net magnetization. A similar second-order phase transition occurs as either the compactness of an individual body or the absolute magnitude of the binding energy of a binary system reaches a large enough value. When this happens, the effective potential of the scalar field changes and a new, spontaneously broken minimum appears. This forces the scalar field to ``roll down'' to a non-zero expectation value.  

The compactness or binding energy at which this phase transition occurs is a function of the coupling constants of the ST theory.
However, the compactness (or the binding energy) of a system, be it an individual NS  or a binary, is also a function of the  NS's equation of state (EoS).
As a result, whether a systems scalarizes or not also depends on the EoS.
We will discuss these theories in more detail in Sec.~\ref{sec:theory}, but it suffices here to say that scalarization (spontaneous, induced or dynamical) only occurs in these theories when a particular coupling constant, $\beta_\ST$, is sufficiently large and negative (for a fixed EoS, NS compactness, and orbital separation). More negative values of this parameter result in scalarization for systems with smaller compactnesses and larger orbital separations~\cite{Damour:1992we,Barausse:2012da,Palenzuela:2013hsa}. 

A little-appreciated problem exists for these theories if one wishes $\beta_{\ST}$ to be negative, such that scalarization can occur. Reference~\cite{Damour:1996ke,PhysRevD.48.3436,PhysRevLett.70.2217} showed that in a cosmological evolution, $\beta_{\ST} > 0$ forces the ST theory to approach GR exponentially, i.e.~GR is an attractor in the theory phase space, and thus the parameterized post-Newtonian (ppN) parameters are exponentially close to their GR values. However, by this same argument, we show in Appendix~\ref{AppA} that
a cosmological evolution with $\beta_{\ST} < 0$ makes GR a repeller, forcing ppN parameters in this theory to deviate from their GR values. 
A more rigorous, nonlinear analysis linking cosmological scales to galactic and eventually Solar System ones would be useful to draw definitive conclusions,
but based on these results, the requirement 
$\beta_{\ST} < 0$, which would enable scalarization, seems incompatible with Solar System experiments.
Regardless, these problems might be avoidable if an external potential (see  e.g. Ref.~\cite{cosmoST}) is included, with a minimum at small values of the scalar field. 

Neglecting the above problems, as done regularly in the 
literature~\cite{Damour:1992we,Damour:1993hw,kramer-wex,Freire:2012mg,
Barausse:2012da,Palenzuela:2013hsa,Shibata:2013pra}, we can study the effects of 
scalarization on astrophysical observations of binary systems, such as binary 
pulsars, and then use these observations to constrain ST theories. If the binary 
components support a scalar field, then an unequal mass binary will decay faster 
than in GR due to the emission of dipolar radiation by the scalar 
field~\cite{Damour:1992we,Damour:1993hw}. Such a decay is stringently 
constrained by binary pulsar observations~\cite{kramer-wex,Freire:2012mg}, which 
can place strong bounds on the existence of dipole radiation, scalarization, and 
the magnitude of $\beta_{\ST}$. Using observations of a pulsar-white dwarf 
binary (J1738+0333) and a single-polytrope EoS with polytropic index $\Gamma 
=2.34$, Ref.~\cite{Freire:2012mg} has constrained $\beta_{\ST} \gtrsim -4.75$. 
Using an APR4 (soft) and an H4 (stiff) EoS, Ref.~\cite{
Shibata:2013pra} used the binary pulsar observations of Ref.~\cite{Freire:2012mg} to constrain $\beta_{\ST} \gtrsim -4.5$ and $\beta_{\ST} \gtrsim -5$ respectively. 

Clearly, constraints on $\beta_{\ST}$ depend on the NS EoS. For a given EoS, only certain sufficiently large, negative values of $\beta_{\ST}$ produce spontaneous scalarization. The bounds quoted above are roughly the least negative values of $\beta_{\ST}$ that allow for scalarization, given a particular EoS. For example, Ref.~\cite{Shibata:2013pra} showed that for the pulsar in Ref.~\cite{Freire:2012mg} with mass $\approx 1.46 M_{\odot}$, an APR4 EoS requires $\beta_{\ST} < -4.5$ for scalarization to occur, while an H4 EoS requires $\beta_{\ST} < -5$, precisely the constraints quoted in Ref.~\cite{Shibata:2013pra}. Thus, binary pulsar observations constrain a region in the $\beta_{\ST}$--EoS space, which means that bounds on $\beta_{\ST}$ can be weakened if one considers stiffer EoSs that lead to less compact stars (larger radius given a fixed mass). Additionally, binary pulsars that have been observed thus far are very widely separated (typical separation larger than $10^{5}$ km), and so their orbital binding energy is not large enough to activate dynamical scalarization. 

The GWs emitted during the late inspiral of NS binaries may also allow for constraints on scalarization, as these waves are produced when the orbital binding energy is very large and, like binary pulsars, GW observations are extremely sensitive to the orbital decay rate. In this paper, we investigate such an idea by 
\begin{enumerate}
\item[(i)] constructing an analytical model of the GWs emitted during late NS inspirals in ST theories, 
\item[(ii)] calculating the response function of interferometric detectors to such waves in the time- and frequency-domains through the stationary phase approximation (SPA), and
\item[(iii)] carrying out a detailed, Bayesian parameter estimation and model-selection study, assuming a GW detection with second-generation detectors (in the currently-planned configuration). 
\end{enumerate}

One of our main results is the following: 
\begin{itemize}[leftmargin=0.2in,rightmargin=0.2in]
\item[] {\bf{GWs emitted in the late inspiral of NS binaries can be used to place constraints on ST theories that are comparable to binary pulsar ones, provided at least one binary component is sufficiently compact to be already spontaneously scalarized by the time the emitted GWs enter the detectors' sensitivity band.}} 
\end{itemize}
Second-generation detectors, like aLIGO, aVirgo, and KAGRA, will be sensitive to GWs emitted by binaries with orbital frequencies above $\sim5$ Hz. If a NS in a binary is sufficiently compact, for a given EoS and value of $\beta_{\ST}$, to be already spontaneously scalarized by the time the binary crosses this frequency threshold, then a GW observation consistent with GR can be used to rule out the existence of dipole radiation and constrain the magnitude of $\beta_{\ST}$. For example, given a polytropic EoS and a NS binary with (gravitational) masses $(1.4074,1.7415) M_{\odot}$, a single GW observation consistent with GR at a signal-to-noise ratio (SNR) of 15 would allow us to place the constraint $\beta_{\ST} \gtrsim -4.5$. This constraint is similar in strength to that obtained with current binary pulsar observations and polytropic EoSs~\cite{Freire:2012mg}, but it is complementary in that it derives from 
sampling the dynamical and non-linear regime of the gravitational interaction. 

Another main result of this paper is the following:
\begin{itemize}[leftmargin=0.2in,rightmargin=0.2in]
\item[] {\bf{Dynamical scalarization in the late inspiral of NS binaries will be difficult to constrain with GWs, unless the system scalarizes at a low enough orbital frequency (i.e.~large enough orbital separation) that a sufficient amount of SNR is accumulated while the ST modifications are active.}}  
\end{itemize}
Reference~\cite{Sampson:2013jpa} proved that an abrupt activation of dipolar 
radiation can only be observed with GWs emitted by binaries if it occurs at an 
orbital frequency well below 50Hz, assuming an aLIGO type detector and an SNR 
of 10. Such a threshold frequency corresponds to an orbital separation larger 
than roughly $35 m$, where $m$ is the total mass of the NS binary, or 
alternatively, $\approx 150$ km for a binary with total mass $3 M_{\odot}$. 
Dynamical scalarization at such large separations only happens for quite specific, finely tuned binary systems 
given 
realistic NS EoSs and values of $\beta_{\ST}$ not already ruled out by binary 
pulsar observations~\cite{Freire:2012mg}. In this analysis, only three systems, 
out of a total of $80$ examined, underwent dynamical scalarization at orbital frequencies 
below $100$ Hz. Of these three, only one exhibited ST effects that were strong 
enough to detect with aLIGO at an SNR of 15. 

We additionally analyze whether another signature of dynamical scalarization --  an early plunge once the scalar field activates -- is detectable with aLIGO-like detectors. We find that:
\begin{itemize}[leftmargin=0.2in,rightmargin=0.2in]
\item[] {\bf{Early plunge and merger due to dynamical scalarization is detectable for certain frequency ranges using model-independent templates.}}  
\end{itemize}
We modeled the early plunge through a \emph{toy model}, a Heaviside truncation of the GW Fourier 
amplitude. For this truncation to be detectable, we find that it must occur between orbital 
frequencies of roughly $45$ Hz and $150$ Hz, assuming an SNR 15 if no truncation was present. 
For plunges that occur at lower frequencies, not enough signal 
would be present in the detector's sensitivity band to lead to a detection in 
the first place. For plunges that occur at higher frequencies, the detector's 
noise is already large enough that such a plunge is difficult to detect. 

The analysis described above required a Markov-Chain Monte-Carlo (MCMC) mapping of the likelihood surface to obtain posteriors and Bayes factors (BFs) between a GR and a non-GR model, given a scalarized GW injection, all of which is computationally expensive. Thus, another important result of this paper is the following:
\begin{itemize}[leftmargin=0.2in,rightmargin=0.2in]
\item[] {\bf{We re-derive a  computationally inexpensive, data analysis measure to determine whether modified gravity effects are detectable, which we call the \emph{the effective cycles of phase}. For the first time, we connect this quantity directly to the Bayes factor. We find that roughly $4$ cycles are needed for a modified gravity effect to be detectable at SNR 15.}}
\end{itemize}
First derived in \cite{PhysRevD.78.124020}, the effective cycles are defined as a certain noise-weighted integral of the GW 
amplitude and the phase difference between a GR and a non-GR inspiral. This 
measure is inspired by the \emph{useful cycles of phase}, introduced in 
Ref.~\cite{Damour:2000gg}, but it differs from this quantity in that the BF can 
be estimated \emph{analytically} in terms of effective 
cycles. Therefore, the effective cycles are directly connected to an important 
data analysis measure of detectability in model hypothesis testing. 

The effective cycles are a noise-weighted dephasing measure that is ideal for studying the distinguishability between models. A non-noise-weighted dephasing is not a good measure, in spite of being often used in theoretical studies, e.g.~\cite{Baiotti:2011am,Yagi:2013sva,Gerosa:2014kta}. The effective cycles take into account the fact that, for departures from GR to be detectable, they must occur in a frequency range in which sufficient SNR is accumulated. This is particularly difficult for GR departures that only become significant at kHz frequencies.

Another question that arose while we carried out this analysis was whether 
custom-made waveforms are necessary to detect the effects of ST gravity, or 
whether model-independent templates are sufficient. This question, of course, is 
SNR dependent, and as explained above, dynamical scalarization is not easily 
measurable with the SNRs expected in advanced detectors, except in a few cases. 
Moreover, it is very difficult to develop 
analytical templates that accurately model the effects of dynamical 
scalarization. We therefore consider the detectability of spontaneous/induced 
scalarization effects with custom-made versus model-independent templates, and 
obtain the following result:
\begin{itemize}[leftmargin=0.2in,rightmargin=0.2in]
\item[] {\bf{More complicated, model-dependent templates are found to be comparably efficient at detecting GR deviations as the simplest, model-independent waveforms one can construct.}}
\end{itemize}
As a proxy for model-independent templates, we use the parameterized post-Einsteinian (ppE) approach of Ref.~\cite{PPE}. The simplest version of such waveforms include only leading post-Newtonian (PN) order modifications to GR in the waveform; thus, they do not contain the precise PN sequence of terms that a custom-made ST theory template (or signal) would possess. This additional structure should allow custom-made templates to be more effective at detecting GR deviations present in the signals they are designed to capture. The enhancement leads to an increased detection efficiency at very high SNR, but negligible effects for signals with SNR's we expect to observe. Additionally, we find that more complicated ppE templates, such as those that include a step-function in the phase to activate a modified gravity effect above a certain frequency, are as effective as the simplest ppE models at detecting ST-type GR deviations. These results are consistent with those in Ref.~\cite{Sampson:2013lpa}.

The rest of this paper presents the details of the results described above and is organized as follows. In Sec.~\ref{sec:theory}, we give an introduction to the non-GR models of interest in this paper. Next, in Sec.~\ref{sec:waveforms}, we describe in detail the methods we used to develop the waveforms tailored to these models. Then, in Secs.~\ref{sec:detect} and~\ref{sec:BF}, we address the question of detectability using effective cycles and a Bayesian analysis respectively. In the final section of this paper, Sec.~\ref{sec:conclusions}, we conclude and point to future research. Throughout this paper, Latin letters refer to spatial indices, Greek letters refer to space-time indices, and we use geometric units in which $G = c = 1$. Additionally, all masses quoted refer to gravitational mass and, unless otherwise specified,
we employ the term ``spontaneously scalarized'' binary (or signal) to
refer to binaries (or signals arising from binaries) whose components undergo spontaneous or induced scalarization. We do so because a necessary condition for induced scalarization to happen is the presence of at least
one spontaneously scalarized star. The term ``dynamically scalarized''  binary (or signal) will
denote binaries (or signals arising from binaries) that undergo dynamical scalarization.

\section{Introduction to Scalar-Tensor Theories \label{sec:theory}} 

In this section, we briefly describe the theory we will study. Initially presented in Ref.~\cite{Damour:1992we,Damour:1993hw}, this theory has recently been revisited in the context of compact binary inspirals and mergers in Ref.~\cite{Barausse:2012da,Palenzuela:2013hsa,Shibata:2013pra}. Here, we review the basics of this theory, following mainly the presentation of Ref.~\cite{Palenzuela:2013hsa}. 

\subsection{Basics  \label{subsec:basics}}
Generic ST theories are defined by the Jordan-frame action
\be
S = \int d^{4}x \frac{\sqrt{-g}}{2 \kappa} \left[ \phi R - \frac{\omega(\phi)}{\phi}  \partial_{\mu} \phi \partial^{\mu} \phi\right] + S_{M}[\chi,g_{\mu\nu}]\,,
\label{eq:Jordan-S}
\ee
where $\kappa = 8 \pi G$, $\phi$ is a scalar field, $R$ is the Ricci scalar associated with the Jordan-frame metric, $g_{\mu \nu}$, and $\chi$ 
are additional matter degrees of freedom that couple directly to the metric. 

The function, $\omega(\phi)$, defines the particular ST theory in play (in some cases there is also a potential function, $V(\phi)$, but here this potential is set to zero). For example, Fierz-Jordan-Brans-Dicke (FJBD) theory~\cite{fierz,jordan,BD} is defined by this action with $\omega(\phi) = \omega_{\BD} = {\rm{const}}$. In this paper, we will consider the class of theories studied in~\cite{Damour:1992we,Damour:1993hw}, which are defined by the action of Eq.~\eqref{eq:Jordan-S} with 
\be
\label{eq:omega-of-phi}
\omega(\phi) = - \frac{3}{2} - \frac{\kappa}{4 \beta \log {\phi}}\,,
\ee
where $\beta$ is a dimensional constant, related to the dimensionless coupling constant of the theory, $\beta_\ST$, by $\beta = (4 \pi G) \, \beta_\ST $. The asymptotic value of $\phi$ at spatial infinity, together with the value of $\beta_\ST$, controls the magnitude of the modifications to GR. 

The Jordan-frame action of Eq.~\eqref{eq:Jordan-S} can be rewritten in the  ``Einstein frame'' via
\begin{multline}
S = \int d^{4}x  \sqrt{-g^{E}} \left(\frac{R^{E}}{2 \kappa} - \frac{1}{2} g^{\mu \nu}_{E} \partial_{\mu} \psi \partial_{\nu} \psi \right) \\+ S_{M}\left[\chi,g_{\mu \nu}^{E}/\phi(\psi)\right]\,,
\end{multline}
where the Einstein-frame metric is related to the Jordan-frame one via $g_{\mu \nu}^{E} = \phi \; g_{\mu \nu}$. The Einstein-frame scalar field $\psi$ is related to its Jordan-frame counterpart via
\be
\left(\frac{d \log \phi}{d\psi}\right)^{2} = \frac{2 \kappa}{3 + 2 \omega(\phi)}\,.
\label{eq:phi-of-psi-DE}
\ee
In the theories of interest in this paper, this differential equation can be solved to obtain
\be\label{eqphi}
\phi = \exp\left(-\beta \psi^{2}\right)\,,
\ee
choosing $\psi=0$ when $\phi = 1$. Using this equation in Eq.~\eqref{eq:phi-of-psi-DE}, one finds
\be
\psi[\omega(\phi)] =  \frac{1}{2 |\beta|} \left[\frac{2 \kappa}{3 + 2 \omega(\phi)} \right]^{1/2}\,.
\ee

In the Einstein frame, the field equations for the metric and the equations of motion for the scalar field are
\begin{align}\label{mod_eins}
G_{\mu \nu}^{E} &= \kappa \left(T_{\mu \nu}^{\psi} + T_{\mu \nu}^{M,E}\right)\,,
\\
\label{eq:scalar-field-eq}
\square^{E} \psi &=- \beta \psi T^{M,E}\,, 
\end{align}
where we have defined
\begin{align}
T_{\mu \nu}^{\psi} &= \partial_{\mu} \psi \partial_{\nu} \psi - \frac{1}{2} g_{\mu \nu}^{E} g^{\alpha \beta}_{E} \partial_{\alpha} \psi \partial_{\beta} \psi\,,
\end{align}
and $T^{M,E}$ is the Einstein-frame trace of $T_{\mu \nu}^{M,E}=T_{\mu \nu}^{M}/\phi$, the matter stress-energy tensor in the Einstein frame.

Because of the field redefinition to Einstein-frame variables, the stress-energy tensor conservation $\nabla_\mu T_M^{\mu\nu}=0$ becomes
\be
\nabla_{\mu}^{E} T^{\mu \nu}_{E} = \beta \psi T_{E} g_{E}^{\mu \nu} \partial_{\mu} \psi\,,
\label{eq:SEP}
\ee
which also follows from the field equations \eqref{mod_eins} and \eqref{eq:scalar-field-eq}. This equation implies, in particular, that test particles (i.e.~point particles with negligible mass) do \textit{not} follow geodesics of the Einstein frame metric $g^E_{\mu\nu}$, although they do follow geodesics of the
original Jordan-frame metric $g_{\mu\nu}$. This means that the weak equivalence principle (i.e., the universality of free fall for
weakly gravitating bodies) is satisfied in these theories. Nevertheless, as will become clearer in the next section, the strong version 
of the equivalence principle (i.e.~the universality of free fall for strongly gravitating bodies) is \textit{not} satisfied in
scalar tensor theories. This is because of the presence of ``scalar charges'' for strongly gravitating bodies, whose appearance can be ultimately traced to
the right-hand side of Eq.~\eqref{eq:SEP} having a non-zero value.

Solar System tracking of the Cassini spacecraft implies the constraint $w_{\BD} > w_{\rm Cassini}\equiv 4 \times 10^{4}$~\cite{will-living,Bertotti:2003rm}
on FJBD theory. Because the class of ST theories that we consider reduces to FJBD in the Solar System (with $\omega_{\BD}=\omega_{0}$ related to the the asymptotic value of $\psi$, which we denote by $\psi_0$, via Eqs.~\eqref{eq:omega-of-phi} and \eqref{eqphi}), this translates into the bound 
\be
\psi_0 \equiv  \frac{1}{2 |\beta|} \left(\frac{2 \kappa}{3 + 2 \omega_{0}} \right)^{1/2} \lesssim 1.26\times 10^{-2} \frac{G^{1/2}}{|\beta|}\,.
\ee
Binary pulsar observations of the orbital decay rate also constrain the theory, since dipolar radiation would greatly accelerate the inspiral~\cite{Damour:1998jk,Freire:2012mg}. These observations require that $\beta_\ST \gtrsim -4.75$~\cite{Freire:2012mg}, but as discussed in the introduction, this constraint depends on which EoS is used~\cite{Shibata:2013pra}.
 
\subsection{Time-Domain Scalar Charges \label{subsec:alphas}}

Given a binary system that consists of two bodies of masses $m_1$ and $m_2$, the evolution of the GWs emitted by this binary will depend on the scalar charges of the two bodies, $\alpha_{1}$ and $\alpha_{2}$. In turn, these charges depend on $\beta_{\ST} \equiv \beta/(4 \pi G)$, but also on the NS compactness $C$ and, in a binary in quasi-circular motion, on the (magnitude of their) orbital velocity $v$ (equivalently, their separation or their orbital frequency). It is therefore necessary to have a good analytic representation of these charges as a function of $\beta_{\ST}$, $C$, and $v$, in order to construct accurate SPA templates. 

The scalar charges are defined by~\cite{Damour:1992we}
\be
\alpha_{A} = -\frac{1}{\sqrt{4 \pi G}} \frac{\partial \ln m_{A}^{E}(\psi)}{\partial {\psi}}\,,
\ee
where $m_{A}(\psi)$ is the mass parameter that enters the point-particle action in the Einstein frame. A related quantity, the {\emph{sensitivity}}, $s_{A}$, can be similarly defined by~\cite{1975ApJ...196L..59E}
\be
s_{A} = \frac{\partial \ln m_{A}(\phi)}{\partial \ln \phi}\,,
\ee
where $m_{A}(\phi)$ is the mass parameter that enters the point-particle action in the Jordan frame. These derivatives are to be taken by keeping the baryonic mass fixed, and the two masses are related via $m_{A}^{E} = m_{A} \phi(\psi)^{-1/2}$. The sensitivities and the scalar charges are then related by~\cite{Damour:1995kt,Mirshekari:2013vb,Palenzuela:2013hsa}
\be
\alpha_{A} = - \frac{2 s_{A} - 1}{\sqrt{3 + 2 \omega_{0}}}\,,
\ee
where $\omega_{0}$ must be greater than $4 \times 10^{4}$ due to the Cassini bound~\cite{will-living,Bertotti:2003rm}. 

In an FJBD theory with a given $\omega_{\rm BD}$, the scalar charges (and thus the sensitivities) are parameters determined exclusively by the compact object's EoSs. Will and Zaglauer~\cite{Will:1989sk} found that in this theory the sensitivities are in the interval $(0.1,0.3)$ for NSs, becoming $0.5$ in the black hole limit. Of course, the scalar charges in this theory are much smaller than the sensitivities, as the former are suppressed by $\approx\omega_{0}^{-1/2}$ relative to the latter. Fixing the EoS, the sensitivities depend only on the mass of the object. Thus, since NS masses are expected to be in the range $(1,2.5) M_{\odot}$, NS binaries have $s_{1} \approx s_{2}$, and dipole radiation is suppressed~\cite{Will:1989sk}. Such suppression explains why it would be difficult to observe or constrain dipole radiation from GWs emitted during binary NS inspirals within FJBD theory.  

In the ST theories of Ref.~\cite{Damour:1992we,Damour:1993hw}, however, the scalar charges can be spontaneously/dynamically generated. In this process of scalarization, the charges can be excited once the gravitational energy of the system exceeds a certain threshold. In isolation, this energy is simply proportional to the NS compactness, $C = M_{ }/R_{ }$, the ratio of the NS mass to its radius. When in a binary, this energy is not only due to the individual compactnesses, but also to the binding energy of the system, which scales as $m_1 m_2/r_{12}$, with $(m_1,m_2)$ the NS masses, and $r_{12}$ the orbital separation. 

\begin{figure}[ht]
\includegraphics[clip=true,angle=-90,width=0.48\textwidth]{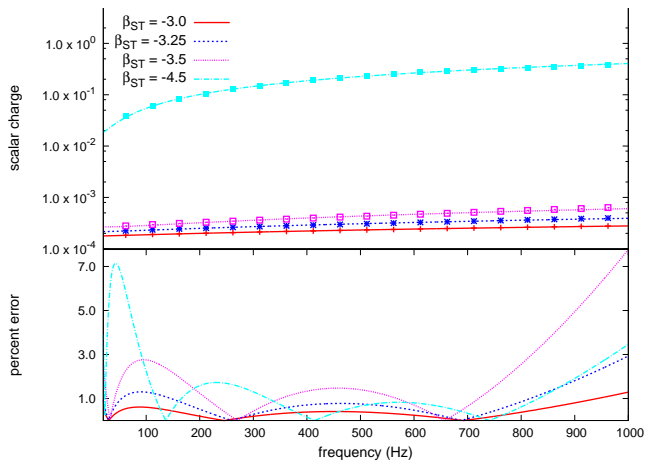} 
\caption{\label{fig:alpha} (Color Online) The upper panel shows the scalar charges for a $1.7415 M_\odot$ NS with a $1.4074 M_\odot$ companion, for various values of $\beta_\ST$ and as a function of GW frequency. The actual data is shown with symbols, and the fits to the data by the different line styles. In the lower panel, we plot the percentage error between the data and the fits. Observe that this error never exceeds $\approx 7\%$.}
\end{figure}

The behavior of the scalar charges during the inspiral and plunge of a NS binary has so far only been calculated semi-analytically in Ref.~\cite{Palenzuela:2013hsa}
for a polytropic EoS with exponent $\Gamma=2$ and maximum NS gravitational mass of $1.8 M_{\odot}$, and these results have been validated by comparing the binary's orbital evolution to the fully non-linear simulations of Ref.~\cite{Barausse:2012da}. Using the results of Ref.~\cite{Palenzuela:2013hsa}, Fig.~\ref{fig:alpha} shows the scalar charge for the more massive star in a NS binary system with $(m_{1},m_{2}) = (1.4074,1.7415) M_{\odot}$ as a function of the dominant GW frequency $f$ (twice the orbital frequency for a quasi-circular binary) up to contact for ST theories with $\beta_\ST = -3.0, -3.25, -3.5$ and $-4.5$. 

As noted, the construction of an SPA waveform in ST theories will require a parameterization of these scalar charges as a function of $\beta_{\ST}$, $m_A$, and $v$.  We desire an analytic expression for the scalar charges because, when calculating the Fourier transform of the GW phase in the SPA, we need to analytically compute {\emph{indefinite}} integrals of functions of these charges. Thus motivated, we will use the following fitting function:  
\be
\label{eq:alpha-fit-poly}
\alpha_{A} =  \sum_{i=0}^{i_{\rm max}} a_{i}^{(A)} \;  v^{i}\,,
\ee
where $v \equiv (\pi m f)^{1/3}$, with $f$ the dominant GW frequency (twice the orbital frequency) and the coefficients $a_{i}^{(A)}$ are functions of $(m_A,m,\beta_\ST)$. We further expand these coefficients as polynomials:
\be a_{i}^{(A)} =\sum_{j=0}^{j_{\rm max}} \sum_{k=0}^{k_{\rm max}} \sum_{\ell=1}^{\ell_{\rm max}} 
(a_{i j k \ell}^{(A)}) \left(-\beta_\ST\right)^{\ell} m_{A}^{j} m^{k}\,
\ee
The more terms kept in the sum, of course, the more closely the function approximates the numerical data. We find empirically that the choice $ i_{\rm max} =  l_{\rm max} = 2$ and $j_{\rm max} = k_{\rm max} = 3$ suffices for our purposes, which leads to $2 \times 3 \times 3 = 18$ fitting coefficients at each PN order.

We use the fitting function described above to fit for the scalar charges as a function of $v$. First, we use the data from Ref.~\cite{Palenzuela:2013hsa} to numerically construct $\alpha_{A}$ as a function of $v$, from a GW frequency of $10$ Hz (the beginning of the aLIGO sensitivity band) up to contact, with a fine discretization, for 37 systems with different masses $(m_{1},m_{2})$ and values of $\beta_\ST$. Each of these data sets is slightly noisy in the low-frequency regime due to small numerical errors, so before proceeding, we smooth each of them with a moving average algorithm using the nearest ten neighbors. 

When carrying out the fits, we do not use the full domain of the data (from $f=10 \; {\rm{Hz}}$ to contact), but rather restrict attention to the low velocity regime. As found in Ref.~\cite{Blanchet:2009sd,Blanchet:2010zd,Blanchet:2010cx}, when fitting numerical data to a PN function, the high velocity regime should not be included in the numerical data. This is because this regime would contaminate the fitting coefficients, as they attempt to capture both the low and high velocity behavior of the function. Moreover, one should not use a very high PN order fitting function, as the high PN order terms would contaminate the low-PN order ones. For these reasons, we choose $i_{\rm max} = 2$ and fit in the region $(10,800) \; {\rm{Hz}}$, which we found empirically to yield robust results, as we will show below. 

\begin{figure}[th]
\includegraphics[clip=true,width=8.5cm]{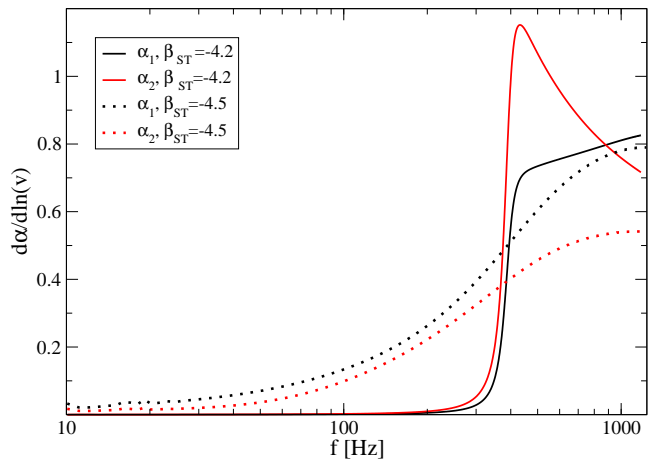} 
\caption{\label{fig:D-scalar} (Color Online) Logarithmic derivative of the scalar charges with respect to velocity as a function of GW frequency for a NS binary with masses $(m_1,m_{2})= (1.4074,1.7415) M_\odot$ computed with the numerical results presented
in Ref.~\cite{Palenzuela:2013hsa}. Observe that the derivative of the charge in the $\beta_\ST = -4.25$ case rises very rapidly around $f = 300 \; {\rm{Hz}}$, rapidly exceeding unity. On the other hand, the derivative in the $\beta_\ST = -4.5$ case remains smaller than unity during the entire inspiral.}
\end{figure}

Finally, in doing the fits, we have neglected scalar charges that undergo dynamical scalarization (i.e.~charges that are close to zero when the binary enters the LIGO band at an orbital frequency of 5 Hz, and then grow very quickly later in the inspiral) for the following reason. The SPA calculation requires that we bivariately Taylor expand all quantities in both $v \ll 1$ \emph{and} in the GR deviation parameter, which in the ST case is $\alpha_{A}$. Although $\alpha_{A} \ll 1$ during the inspiral, $d\alpha_{A}/d \ln{v} > 1$ at a certain frequency, around $f = 300  \; {\rm{Hz}}$ for the cases considered in Fig.~\ref{fig:D-scalar}. Thus, during dynamical scalarization, one should not expand in derivatives of 
$\alpha_{A}$, which makes any analytic treatment very difficult. Third, a polynomial in velocity is ill-suited for representing scalar charges that undergo dynamical scalarization. If one insisted on using such a fitting function, then the coefficients $a_{i}^{(A)}$ would grow by factors of $10^{3}$ with increasing $i$. Such a highly divergent behavior of the fitting coefficients renders any subsequent PN expansion useless. 

Given this, we fit Eq.~\eqref{eq:alpha-fit-poly} independently to each set of clean data corresponding to a \emph{particular} value of $\beta_\ST$  rather than fitting all the data with the same set of coefficients at the same time. We proceed this
way because the scalar charges for larger values of $|\beta_\ST|$ (i.e., $\beta_\ST = -4.5$) are much larger than, for example, the charges at $\beta_\ST=-3.5$. This, paired with our lack of data for values of $|\beta_{\ST}| < 3.0$ and the sparseness of the data in $(m_{1},m_{2})$ space, causes our algorithm to generate poor fits for small values of $\beta_\ST$. In particular, the fitting function obtained by simultaneously fitting all the data does not  monotonically approach zero as $\beta_\ST \to 0$. We therefore arrive at a set of $n=18$ coefficients at each PN order for each different value of $\beta_\ST$. Figure~\ref{fig:alpha} shows the fitted function and the numerical data for several values of $\beta_\ST$, where note the error in the fit never exceeds $7 \%$ in the frequency window $[10:10^{3}]$ Hz.

\section{Developing Waveforms for Scalar-Tensor Theories \label{sec:waveforms}}

In this section, we describe how to construct the time-domain response of the aLIGO detectors to impinging GWs. We use the restricted PN approximation, following mostly the presentation in Ref.~\cite{Yunes:2009yz,Chatziioannou:2012rf}. We then construct the frequency domain, SPA waveforms from these time-domain functions, and discuss their regimes of validity.

\subsection{Time-Domain Response}
\label{sec:time-domain-response}

The response of a GW interferometer to a signal can be computed by investigating how the metric perturbation affects the geodesic deviation. Doing so, one finds~\cite{Kidder:1995zr,poisson2014gravity}
\be
\label{eq:t-dom-resp}
h(t) = F_{+} h_{+} + F_{\times} h_{\times} + F_{\br} h_{\br} + F_{\Lo} h_{\Lo} + F_{\se} h_{\se} + F_{\sn} h_{\sn}\,. 
\ee
The quantities $(F_{+},F_{\times},F_{\br},F_{\Lo},F_{\se},F_{\sn})$ are beam or angular pattern functions that depend on the geometry of the detector [see e.g.~Eqs.(2)--(7) in Ref.~\cite{Chatziioannou:2012rf}]. The quantities $(h_{+},h_{\times},h_{\br},h_{\Lo},h_{\se},h_{\sn})$ are the six possible GW polarizations in a generic theory of gravity: the plus-mode, the cross-mode, the scalar ``breathing'' mode, the scalar ``longitudinal'' mode, and the two vector modes.
Recall that GR, being a massless spin-2 theory, has only two propagating degreees of freedom.

The waveform polarizations can be computed from the trace-reversed metric perturbation in the far-zone (at a distance much greater than a GW wavelength from the center of mass of the binary) via
\begin{align}
h_{+} &= e^{+}_{ij} \left( P^{i}_{m} P^{j}_{l} \bar{h}^{ml} - \frac{1}{2} P^{ij} P_{ml} \bar{h}^{ml} \right)\,,
\\
h_{\times} &= e^{\times}_{ij} \left( P^{i}_{m} P^{j}_{l} \bar{h}^{ml} - \frac{1}{2} P^{ij} P_{ml} \bar{h}^{ml}\right)\,,
\\
h_{\br} &= \frac{1}{2} \left(\hat{N}_{jk} \bar{h}^{jk} - \bar{h}^{00} \right)\,,
\\
h_{\Lo} &= \hat{N}_{jk} \bar{h}^{jk} + \bar{h}^{00} - 2 \hat{N}_{j} \bar{h}^{0j}\,,
\\
h_{\sn} &= e_{k}^{x} \left[P^{k}_{j} \left(\hat{N}_{i} \bar{h}^{ij} - \bar{h}^{0j} \right)\right]\,,
\\
h_{\se} &= e_{k}^{y} \left[P^{k}_{j} \left(\hat{N}_{i} \bar{h}^{ij} - \bar{h}^{0j} \right)\right]\,,
\end{align}
where we use multi-index notation $\hat{N}^{l_1 \ldots l_n} = \hat{N}^{l_1} \cdots 
\hat{N}^{l_n}$. Here, $\hat{N}^{i}$ is a unit vector pointing from the source 
to the detector, $P_{ij} = \delta_{ij} - \hat{N}_{ij}$ is a projection operator 
orthogonal to $\hat{N}^{i}$,  and $e_{i}^{x}$ and $e^{y}_{i}$ are basis vectors
orthogonal to $\hat{N}^{i}$. If we choose a coordinate system defined by the 
triad $(\hat{i}^{i},\hat{j}^{i},\hat{k}^{i}$), such that $\hat{N}^{i}$ is given 
by
\be
\hat{N}^{i} = \sin{\iota} \; \hat{j}^{i} + \cos{\iota} \; \hat{k}^{i}\,,
\ee
then 
\begin{align}
e_{x}^{i} &= - \hat{i}^{i}\,,
\\
e_{y}^{i} &= \cos{\iota} \; \hat{j}^{i} - \sin{\iota} \; \hat{k}^{i}\,,
\\
e^{+}_{ij} &= \frac{1}{2} \left(e^{y}_{i} e^{y}_{j} - e^{x}_{i} e^{x}_{j}\right)\,,
\\
e^{+}_{ij} &= \frac{1}{2} \left(e^{x}_{i} e^{y}_{j} + e^{y}_{i} e^{x}_{j}\right)\,,
\end{align}
where $\iota$ is the inclination angle, such that $\hat{L}^{i} \hat{N}_{i} = \cos{\iota}$, with $\hat{L}^{i}$ the unit orbital angular momentum vector.

The ST theories of Ref.~\cite{Damour:1992we,Damour:1993hw} have, in principle, three propagating GW modes: the two transverse-traceless modes ($h_{+}$ and $h_{\times}$) and the breathing mode ($h_{\br}$). In the Eardley, {\emph{et al}} classification~\cite{Eardleyprd}, these modes correspond to the excitation of the Newman-Penrose scalars $\Psi_{4}$ and $\Phi_{22}$. One can show~\cite{Barausse:2012da}, however, that $\Phi_{22}$ is proportional to $\psi_{0}$ (or alternatively to $\omega_{0}^{-1/2}$) to leading order in a $\psi_{0} \ll 1$ expansion; $\Psi_{4}$ is, of course, independent of $\psi_{0}$ to leading order. Thus, the effect of $h_{\br}$ in the response $h(t)$ is subleading in $\psi_{0}$ relative to the effect of $h_{+}$ and $h_{\times}$~\cite{Barausse:2012da}. In fact, this is exactly the same as in standard FJBD theory~\cite{TEGP,Will:1994fb}. Even if this were not the case and the response to the breathing mode were not suppressed as $1/\sqrt{\omega_{0}}$, the detectability of this mode would require 
a network of detectors~\cite{Nishizawa:2009bf,Chatziioannou:2012rf,Hayama:2012au,Nishizawa:2013eqa,Horava:2014mra}. Given this, we will neglect the breathing mode in the response function.

\subsection{Time-Domain Waveform \\ and the Restricted PN Approximation}

As explained above, the interferometer response is dominated by the plus- and cross-polarized metric perturbations, $h_{+}(t)$ and $h_{\times}(t)$, which must be obtained by solving the modified field equations in the PN approximation. In the far-zone, these waves can be written as the following PN sum:
\be
\label{eq:def-h+x}
h_{+,\times}(t) = \frac{2 G \eta m}{D_{L}} x(t) \sum_{p=0}^{+\infty} x(t)^{p/2} H_{+,\times}^{(p/2)}(t)\,,
\ee
where $D_{L}$ is the (luminosity) distance from the source to the detector, $\eta = m_{1} m_{2}/m^{2}$ is the symmetric mass ratio, and recall that $m= m_{1} + m_{2}$ is the total mass, and $x(t) = [2 \pi G_{\rm eff} m F(t)]^{2/3}$ is the (time-dependent) PN expansion parameter of leading ${\cal{O}}(1/c^{2})$, with $F(t)$ the orbital frequency. Notice that $x$ depends on $G_{\rm eff} \equiv G (1 + \alpha_{1} \alpha_{2})$, because in these theories the Newtonian force of attraction between two bodies, and thus, Kepler's third law of orbital motion, takes on the usual Newtonian expression, but with the replacement $G \to G_{\rm eff}$~\cite{Damour:1992we,Will:1989sk}. Recall that $\alpha_{1}$ and $\alpha_{2}$ are the scalar charges of NSs, which depend on the internal structure of the bodies and on the gravitational binding energy of the system, as discussed in Sec.~\ref{subsec:alphas}. 

The time-functions $H_{+,\times}^{(p/2)}$ can always be written in terms of an amplitude ($A_{+,\times}^{(p/2,n)}(t)$ or $B_{+,\times}^{(p/2,n)}$(t)) and the binary's orbital phase $\phi(t)$:
\be
H_{+,\times}^{(p/2)} = \sum_{n=0}^{\infty} \left(A_{+,\times}^{(p/2,n)} \cos{n \phi} + B_{+,\times}^{(p/2,n)} \sin{n \phi} \right)\,.
\ee
To leading PN order (keeping only the $p=0$ term), the oscillating part of Eq.~\eqref{eq:def-h+x} reduces to
\begin{align}
\label{eq:restrictedPN-hp}
h_{+}(t) &= 2 {\cal{A}}(t)   \left(1 + \cos^{2}{\iota}\right) \cos{2 \phi(t)}\,, 
\\
\label{eq:restrictedPN-hc}
h_{\times}(t) &= -4 {\cal{A}}(t) \cos{\iota} \sin{2 \phi(t)}\,.
\end{align}
Here we have defined the time-dependent amplitude
\be
{\cal{A}}(t) = - \frac{{\cal{M}}}{D_{L}} [2 \pi G m F(t)]^{2/3}\,,
\ee
with ${\cal{M}} = \eta^{3/5} m$ the chirp mass, and $\iota$ is the inclination. Notice that this is functionally exactly the same result as in GR, because, as discussed in Sec.~\ref{sec:time-domain-response}, ST corrections to the amplitude scale as $\psi_{0}$, and are thus subleading. 

The \emph{restricted PN approximation} consists of approximating the 
time-domain waveform by the leading PN order terms in the waveform amplitudes, 
without restricting the PN order in the waveform phase. That is, one keeps only the 
leading, $p=0$ term in Eq.~\eqref{eq:def-h+x}, thus obtaining 
Eqs.~\eqref{eq:restrictedPN-hp} and~\eqref{eq:restrictedPN-hc}, but as many PN 
terms as one wishes in the orbital phase $\phi(t)$. Such an 
approximation is reasonable in a first analysis because interferometric 
detectors are much more sensitive to the phase of the response than the 
amplitude.  

The plus- and cross-polarized metric perturbations in ST theories, and thus the time-domain interferometric response, are different from those predicted by GR mainly because of the temporal evolution of the orbital phase. The phase can be obtained by integrating the expression 
\be
\label{eq:Fdot}
\ddot{\phi} = 2\pi \dot{F} = 2 \pi \frac{dE_{b}}{dt} 
\left(\frac{dE_{b}}{dF}\right)^{-1}
\ee
twice, where $E_{b}$ is the gravitational binding energy.
By the balance law, $d E_{b}/dt$ must equal (minus) the luminosity ${\cal{L}}$ of all propagating degrees of freedom in the far-zone. In GR, the only energy loss is due to the emission of GWs, but in ST theories one must also account for dipolar radiation induced by the propagation of the scalar mode: 
\be
\label{eq-luminosity}
{\cal{L}} = \frac{G}{3} \eta^{2} \left(\alpha_{1} - \alpha_{2}\right)^{2} x^{4} + \frac{32}{5} G \eta^{2} x^{5}\,,
\ee
where the first term is due to dipole radiation and the second due to quadrupolar radiation. We then see that ST theories modify the evolution of the orbital phase, which is precisely the component of the interferometric response that detectors are most sensitive to.

\subsection{Fourier Response \\ and the Stationary Phase Approximation}

In GW data analysis, one uses the Fourier transform of the response function 
(the Fourier response), instead of the time-domain response. 
In computing the frequency- or time-domain response, 
one can effectively neglect the time-dependence of 
the beam pattern functions. These are induced by the rotation and motion of the 
Earth, which occurs on a time-scale much longer than the duration of the GW 
signal, which is typically less than 20 minutes for the signals that fall in the sensitivity band of 
second-generation, ground-based detectors, such as aLIGO,  aVirgo, and KAGRA.
Thus, the Fourier response can be written 
as 
\be
\tilde{h}(f) = \int_{-\infty}^{\infty} dt \, e^{2 \pi i f t} \left[F_{+} h_{+}(t) + F_{\times} h_{\times}(t) \right]\,,
\label{eq:F-response-def}
\ee
where $(F_{+},F_{\times})$ are effectively constant. 

Using the restricted PN approximation [Eqs.~\eqref{eq:restrictedPN-hp} and~\eqref{eq:restrictedPN-hc}], we can rewrite the Fourier response of Eq.~\eqref{eq:F-response-def} as
\be
\tilde{h}(f) = \int_{-\infty}^{\infty} dt \, e^{2 \pi i f t} {\cal{A}}(t)  \left[Q_{C} \cos 2 \phi(t) + Q_{S} \sin 2 \phi(t)\right]\,,
\label{eq:F-response-new}
\ee
where we have defined the cosine- and sine-projected beam-pattern functions 
\begin{align}
Q_{C} &= 2 \left(1 + \cos^{2}{\iota}\right) \cos{2 \Psi} F_{+} - 4 \cos{\iota} \sin{2 \Psi} F_{\times}\,,
\\
Q_{S} &= 2 \left(1 + \cos^{2}{\iota}\right) \sin{2 \Psi} F_{+} + 4 \cos{\iota} \cos{2 \Psi} F_{\times}\,,
\end{align}
with $\Psi$ the polarization angle. For binary systems that are in a quasi-circular orbit and whose binary components are not spinning, the inclination angle and $Q_{C,S}$ are all constant.  The quasi-circular and non-spinning approximations are sufficient for our analysis because most NS binaries are expected to have circularized and spun-down by the time they enter the sensitivity band of second-generation, ground-based detectors.  

The integral in Eq.~\eqref{eq:F-response-new} that defines the Fourier response falls in the class of generalized Fourier integrals. When the integrands have a stationary point, namely a time $t_{0}$ at which $\dot{\phi}(t_{0}) = \pi f$, the integral can be approximated via the method of steepest descent, which to leading order reduces to the SPA~\cite{Bender,Yunes:2009yz}. For this approximation to hold, the amplitude of the integrand must be slowly-varying, while the phase must be rapidly oscillating, such that the integral is non-vanishing only in a small neighborhood around the stationary point. 

Within the SPA, the integral in Eq.~\eqref{eq:F-response-new} reduces to
\be
\tilde{h}(f) = \tilde{A} f^{-7/6} e^{i \Psi(f)}\,,
\ee
where $f$ is the Fourier or GW frequency (twice the orbital frequency), the Fourier amplitude is a constant given by
\be
\tilde{A} \equiv - \left(\frac{5}{384} \right)^{1/2} \pi^{-2/3} \frac{{\cal{M}}^{5/6}}{D_{L}} \left(Q_{C} + i Q_{S}\right)\,,
\ee
and the Fourier phase must be computed from~\cite{cutlerflanagan,Yunes:2009yz}
\be
\label{eq:SPA-phase-def}
\Psi(f) = 2 \pi \int^{f/2} dF' \left(2 - \frac{f}{F'}\right) \frac{F'}{\dot{F}(F')}\,.
\ee
This expression makes it clear that the dominant ST modifications to the Fourier response are due to modifications to the rate of change of the orbital frequency.

\subsection{SPA Templates in Scalar-Tensor Theory \label{sec:SPAtemp}}

We now follow the analysis of the previous subsection and provide explicit formulas for the Fourier transform in the SPA of restricted PN waveforms in ST theories to 1PN order. We first focus on the binding energy and its rate of change. The former is presented in Eq.~$(6.4)$ of Ref.~\cite{Mirshekari:2013vb} in terms of the individual masses and velocities of the system. One can transform the binding energy to relative coordinates, through the mapping in Eqs.~$(6.9)$--$(6.11)$ of the same paper. The rate of change of the orbital binding energy is equal to the energy flux carried by all propagating degrees of freedom by the balance law. This quantity is presented in Eqs.~$(6.16)$ and $(6.17)$ of Ref.~\cite{Mirshekari:2013vb} in relative coordinates. 

The ST theory-modified version of Kepler's third law of motion can be computed by solving $|a^{i}| = r_{12} \omega^2$ in a PN expansion for $r_{12}$, the relative orbital separation, using also that $|v_{12}^{i}| = r_{12} \omega$, where recall that we are considering only binaries in quasi-circular orbits. The relative acceleration $a^{i}$ is given in Eq.~$(1.4)$ and $(1.5)$ of Ref.~\cite{Mirshekari:2013vb}, where $\omega$ is the orbital angular velocity. We find that
\begin{align}
r_{12} &= \frac{m}{x} \left\{\left(1 + \frac{1}{3} \alpha_{1} \alpha_{2}\right) + x \left[\left(\frac{\eta}{3} - 1\right) 
\right. \right. 
\nn \\
&\left. \left.
+ \frac{1}{3} \alpha_{1} \alpha_{2}  \left(\eta - 1\right)\right]  + {\cal{O}}(x^{2},\alpha_{A}^{4}) \right\} \,,
\end{align}
where we have set $\psi_{0} = 0$, as this quantity is constrained to be less than $10^{-2}$ by current observations. 

With the above modified version of Kepler's third law, one can rewrite the binding energy and its rate of change as a function of the PN expansion parameter. Using the definition of Eq.~\eqref{eq:Fdot}, the rate of change of the orbital frequency is then
\begin{align}
\dot{F} &= \frac{1}{2} \frac{\eta}{\pi m^{2}} x^{9/2} \left(\alpha_{1} - \alpha_{2}\right)^{2} 
\nn \\ 
&+ \frac{48}{5} \frac{\eta}{\pi m^{2}} x^{11/2} \left\{1 + \frac{1}{576} \left[ \left(-15 - 35 \eta\right) \alpha_{1}^{2} 
\right. \right. 
\nn \\
&\left. \left.
+ \frac{1}{576} \left(35 \eta + 63 \right) \alpha_{1} \alpha_{2} + 
\alpha_{1} \alpha_{1}'  \left( \frac{5}{144} \nu - \frac{5}{48} \right)
\right. \right. 
\nn \\
&\left. \left.
- \alpha_{1} \alpha_{2}' \left( \frac{5}{144} \nu + \frac{43}{48}\right)
+ 1 \to 2 \right] \right\} + {\cal{O}}(x^{6},\alpha_{A}^{4})\,,
\end{align}
where $\alpha_{A}' := d\alpha_{A}/d\ln{v}$. 
The first term in this expression corresponds to dipolar radiation, while the second one is the usual quadrupolar radiation term of GR. Notice that both are corrected by terms proportional to the scalar charges, due to the modification to Kepler's third law. 

With all of this at hand, we can now compute the Fourier phase in the SPA through Eq.~\eqref{eq:SPA-phase-def}. In Sec.~\ref{subsec:alphas}, we fitted $\alpha_A$ through Eq.~\eqref{eq:alpha-fit-poly}, which can be rewritten as 
\be
\alpha_A = c_A + d_A v + \ldots,
\ee
identifying $a_{0}^{(A)}$ with $c_{A}$ and $a_{1}^{(A)}$ with $d_{A}$, which are themselves functions of $m_A$, $m$, and $\beta_\ST$.  Finally, we find
\begin{align}
\Psi(f) &= 2 \pi f t_{c} - \phi_{c} - \frac{\pi}{4}  + \frac{3}{128 \eta v^{5}} 
\left[- \frac{5}{168} \left(c_{1} - c_{2}\right)^{2} v^{-2}   
\right.
\nonumber \\
&- \left. \frac{8400}{109489} (c_1 - c_2)(d_1 - d_2)v^{-1} + \ldots \right]\, .
\label{eq:SPAphase}
\end{align}
We see that the main modifications to the Fourier response in the SPA is due to the scalar charges, which induce a dipole correction to leading order. We have checked that the dipole term is exactly what one obtains in FJBD theory, when one rewrites $c_{1,2}$ in terms of the sensitivities $s_{1,2}$. 

The ST-modified Fourier response described above is only valid for systems of \emph{unequal} mass. In the case of equal-mass binaries, the $-1$PN and the $-0.5$PN terms in Eq.~\eqref{eq:SPAphase} vanish, because then $s_{1} = s_{2}$. In such a system, the lowest PN order correction to the Fourier phase enters at $0$PN order. Henceforth, we only use the ST-modified SPA for analysis of binaries whose component masses are unequal.

\subsection{Comparison of SPA Phase to Numerical Phase \label{sec:comp}}

We can now validate the SPA model of the previous subsection by comparing it to the Fourier transform of the time-domain numerical solutions for the GWs presented in Ref.~\cite{Palenzuela:2013hsa}. Before we do so, we define the measure we will use for such a validation: the dephasing $\Delta \Psi$. This quantity is defined as 
\be
\label{eq:dephasing}
\Delta \Psi_i = {\rm{min}}_{t_{c},\phi_{c}} \left| \Psi_{\GR} - \Psi_{\ST}\right|\,,
\ee
where $\Psi_{\GR}$ is the Fourier phase in GR, $\Psi_{\ST}$ is the 
Fourier phase in ST theories, and the subindex $i$ labels four different strategies we employ to evaluate $\Psi_{\GR}$ and $\Psi_{\ST}$ as described below. 
This measure requires minimization over time and phase of 
coalescence, i.e.~over a constant time and phase shift. Such a 
minimization is required to compare different template families,
e.g.~a time-domain waveform to a frequency-domain one.

In order to validate the SPA model of the previous subsection, we will evaluate the dephasing of Eq.~\eqref{eq:dephasing} in four different ways: 
\begin{enumerate}
\item {\bf{Numerical}}: $\Psi_{\GR}$ and $\Psi_{\ST}$ are given by the Fourier transforms of the time-domain GW data of Ref.~\cite{Palenzuela:2013hsa}, minimizing the difference over phase and time of coalescence.
\item {\bf{-1 PN}}: $\Psi_{\ST}$ is given by Eq.~\eqref{eq:SPAphase}, keeping only the leading PN term (the $-1$PN term).
\item {\bf{-0.5 PN}}: $\Psi_{\ST}$ is given by Eq.~\eqref{eq:SPAphase}, keeping the leading and first subleading PN terms (the $-1$ and the $-0.5$PN terms).
\item {\bf{0 PN}}: $\Psi_{\ST}$ is given by Eq.~\eqref{eq:SPAphase}, keeping terms up to Newtonian order (the $-1$, $-0.5$ and $0$PN terms). 
\end{enumerate}
In all $N$PN cases, $\Psi_{\GR}$ is the GR Fourier phase in the SPA, for example given in Ref.~\cite{Blanchet:2001ax,PPE}. By comparing the numerical dephasing to the $N$PN ones, we will validate the SPA templates constructed in the previous subsection. 

\begin{figure}[ht]
\includegraphics[clip=true,width=0.45\textwidth]{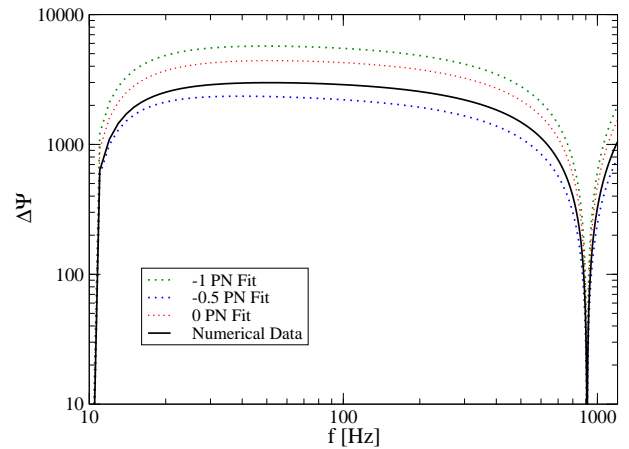} 
\caption{\label{fig:dephasing} (Color Online) Dephasing between a GR and ST signal as a function of GW frequency, calculated using both the numerical Fourier phases (solid) and the SPA Fourier phases (dotted curves) for a binary with masses $(1.4074,1.7415) M_{\odot}$ and $\beta_{\ST} = -4.5$. The ST SPA phases were calculated to various PN orders. }
\end{figure}

Such a comparison is carried out in Fig.~\ref{fig:dephasing}. The solid line is 
the numerical dephasing, while the dotted lines are the $N$PN dephasings in the 
SPA. Observe that the SPA dephasing at $-0.5$PN order is a better approximation than keeping the dephasing to $0$PN
or $-1$PN order, relative to the numerical dephasing. 
We will therefore truncate the fit to the scalar charges at $-0.5$PN 
order for the rest of this paper (i.e, we keep only two ST PN corrections) and model the ST SPA templates by Eq.~\eqref{eq:SPAphase}, 
without ST higher-order PN terms. 

\subsection{ppE Waveforms}
Before proceeding with a detailed data analysis study, we map the SPA templates of Sec.~\ref{sec:SPAtemp} to the ppE templates constructed in Ref.~\cite{PPE}. The ppE family is a theory-independent set of templates, designed to capture model-independent deviations from GR with GW observations. This family is constructed by introducing parameterized modifications to both the binding energy and the energy balance equations of GR. Both types of modifications lead to changes in the GW phase, which can be described by introducing a small number of new parameters into the GW template.

The full GW waveform for the coalescence of two compact objects is typically split into three sections: inspiral, merger, and ringdown. ppE templates have been developed for all three phases, but in this paper we are interested only in the inspiral portion. The latter can be defined as the part of the waveform that is generated before the two bodies plunge into each other. The definition of the end of inspiral is somewhat arbitrary, but we follow typical conventions and define the transition from inspiral to merger as occurring at the innermost stable circular orbit of the system in center of mass coordinates (or alternatively at contact). In any case, the merger of NS binaries occurs at the very high frequency end of aLIGO's sensitivity band, almost outside of it altogether.

The simplest, quadrupole ppE inspiral templates have the form
\be
\tilde{h} (f) = \tilde{h}^{\GR}\cdot (1+\alpha_{\ppE}  u^{a}) e^{i\beta_{\ppE} u^{b}}, \quad\quad u = (\pi \mathcal{M} f)^{1/3},
\label{eq:ppEtemp}
\ee
where $\tilde{h}^{\GR}$ is the Fourier response in GR. These simple ppE waveforms contain an additional amplitude term, $\alpha_{\ppE} u^{a}$, and an additional phase term, $\beta_{\ppE}  u^{b}$, relative to GR. We refer to $\alpha_{\ppE} $ and $\beta_{\ppE} $ as the \emph{strength} parameters of the ppE deviations, and to $a$ and $b$ as the \emph{exponent} parameters. 

These ppE waveforms cover all known inspiral waveforms from specific alternative theories of gravity \cite{cornish-PPE} that are analytic in the frequency evolution of the GWs. Some specific examples are discussed in Ref.~\cite{Yunes:2013dva}. They can not, however, perfectly match the signals generated by the theories of interest in this paper. For the case of induced/spontaneous scalarization, Sec.~\ref{sec:comp} showed that one needs \emph{two} ST PN corrections to the SPA phase (i.e. a $-1$PN and a $-0.5$PN term), while the simplest ppE model includes only one. For the dynamically scalarized case, the corrections to the Fourier response are very abrupt (almost a step function), which cannot be captured with a single PN term. In spite of this, we will see that the ppE model is quite adequate at detecting a ST deviation, provided the deviation is strong enough to be detectable in the first case.

\section{Detectability of Scalar-Tensor Deviations Through an Effective Cycles Approach \label{sec:detect}}

In this section, we carry out the first of a two-part data analysis investigation to determine the detectability of ST deviations in GWs emitted during the late inspiral of NS binaries. We first construct a new, computationally inexpensive data analysis measure to determine when a GR deviation is sufficiently loud for detection with aLIGO-type detectors. We then use this measure on ST signals and ppE signals to estimate their detectability.

\subsection{Useful and Effective Cycles of Phase \label{sec:usecyc}}

Model hypothesis testing, i.e.~the determination of whether model A or B is better supported by some data, usually requires a detailed Bayesian analysis through MCMC techniques that map the likelihood surface and the posterior distributions of template parameters in order to calculate BFs\footnote{The BF, assuming equal priors for the two competing theories, is the odds that one theory is favored by the data over another theory. For instance, a BF of 100 in favor of GR means that there is a 100:1 ``betting odds'' that GR is the correct theory given the data. In this paper, we are considering only nested models. For example, GR is recovered from ppE templates when the strength parameter $\beta_\ppE = 0$. In this case, the BF can be calculated from the Savage-Dickey density ratio, which compares the prior weight at $\beta_\ppE =0$ to the posterior weight at that value. The BF is then calculated via $BF = p(\beta_\ppE = 0 | d)/p(\beta_\ppE = 0)$. If there is more posterior weight at this point than prior weight, 
the model selection process favors GR.}. Such studies are computationally expensive, and it is therefore desirable to construct a simple and computationally inexpensive measure for accomplishing similar goals. The construction of this measure is the topic of this section.

We first describe a quantity that has been used in the literature as a stand-in for the importance of a particular GW phase term~\cite{Damour:2000gg}: the \emph{useful cycles}, $\mathcal{N}_u$. This quantity is defined in \cite{Damour:2000gg} as
\begin{multline}
\mathcal{N}_u = \left(\int_{F_\fmin} ^{F_{\tiny \mbox{max}}}  d\ln f \frac{a^2(f)}{S_n(f)} \frac{d\phi}{2\pi df}\right) \\\times \left( \int_{F_\fmin} ^{F_{\tiny \mbox{max}}} d\ln f \frac{a^2(f)}{f S_n(f)}\right)^{-1},
\label{Eq:usecyc}
\end{multline}
which is essentially a noise-weighted measure of the total number of cycles of phase due to any particular term in the phase evolution. In Eq.~\eqref{Eq:usecyc}, $a(f)$ is defined by $|\tilde{h}(f)|^2 = \tilde{A}^2(f) =N(f) a^2(f)/f^2$,
with $N(f) = (1/2\pi)(d\phi/d\ln F)= F^2/(dF/dt)$. The expression for $\mathcal{N}_u$ can be re-expressed in terms of the characteristic strain $h_c(f) = \sqrt{f} \tilde{A}(f)$ as
\be
\mathcal{N}_u = {\rm SNR}^2 \left( \int_{F_\fmin} ^{F_{\tiny \mbox{max}}}  \frac{h_c^2(f)}{S_n(f)} \frac{1}{N(f)}\; d\ln f \right)^{-1}.
\label{Eq:usecyc2}
\ee 
Thus, the number of useful cycles is equal to the harmonic mean of $N(f)$, with a weighting factor equal to the SNR squared per logarithmic frequency interval, $\Delta {\rm SNR}^2(f)=h_c^2(f)/S_n(f)$. 

The difference in the number of useful cycles between waveform models is sometimes used as a proxy for the detectability of the difference in the models (see eg. Ref.~\cite{Baiotti:2011am,Yagi:2013sva,Gerosa:2014kta}). One must be very careful when doing this for two reasons. The first is made clear by re-writing Eq.~\eqref{Eq:usecyc} in the form of Eq.~\eqref{Eq:usecyc2}.  This re-casting of the useful cycles shows that it is not permissible to simply replace $N(f)$ with $\Delta N(f)$, where $\Delta N(f)$ is the change that is introduced by a particular modification to the phase. In order to calculate $\mathcal{N}_u$ due to a change in the phase evolution, it is necessary to calculate both $\mathcal{N}_u$ from the original phase and from the changed phase, and then take the difference. This is not a problem, \emph{per se} - it is simply an issue that must be kept in mind when calculating $\mathcal{N}_u$ for a particular phase term.

A larger issue with $\mathcal{N}_u$ as a measure of detectability is the murkiness of its connection with quantities such as the BF, which are directly related to model selection. The logarithm of the BF, as derived in Ref.~\cite{cornish-PPE}, satisfies
\be
\log {\rm BF} \sim \frac{1}{2}(1-{\rm FF}^2) {\rm SNR^2} + {\cal{O}}[(1 - {\rm{FF}}^{2})^{2}]\,.
\ee
We can use the following expression for the fitting factor, ${\rm FF}$, given two waveforms with the same amplitude $\tilde{A}(f)$, but with phases that differ by $\Delta \Psi(f)$:
\be
{\rm FF} ={\rm SNR^{-2}} \, \max_{\lambda^{a}} \left(\int \frac{h_c^2(f) \cos(\Delta\Psi(f))}{S_n(f)} \; d\ln f\right)\,,
\ee
where the maximization is done over all system parameters $\lambda^{a}$.  In the limit ${\rm FF} \sim 1$, i.e.~for small deviations from GR, these expressions can be combined to give
\begin{align}
\log {\rm BF} &\sim \frac{1}{2} \min_{\lambda^{a}} \int \frac{h_c^2(f) \Delta\Psi^2(f)}{S_n(f)} 
d\ln f  + {\cal{O}}(\Delta \Psi^{4})\;.
\label{Eq:bf}
\end{align}

Given the above expression for the BF, a natural definition for a computationally inexpensive data analysis measure presents itself, the \emph{effective cycles of phase}:
\be
\mathcal{N}_e = \min_{\Delta t, \Delta \phi} \left[\frac {1}{2\pi{\rm SNR}}  
\left( \int \frac{h_c^2(f) \Delta \Phi^2(f)}{S_n(f)} \; d\ln f 
\right)^{1/2}\right]\,,
\label{Eq:EffCyc}
\ee
where $\Delta \Phi \equiv \Delta \Psi(f) + 2 \pi f \Delta t - \Delta \phi$, 
where $\Delta t$ and $\Delta \phi$ are an arbitrary time 
and phase shift respectively. As before, the dephasing is
$\Delta \Psi = \Psi_{1}(f) - \Psi_{2}(f)$, where in our case $\Psi_{1}$ will be 
the Fourier phase of a GR signal
and $\Psi_{2}$ the phase of a non-GR signal, e.g.~for a ppE waveform, $\Delta 
\Psi(f) = \beta_\ppE u^b$.
We define ${\cal{N}}_{e}$ with a $(\Delta t, \Delta \phi)$-minimization  
because the non-GR terms induce a modification in the frequency and phase 
evolutions that renders meaningless a direct comparison between time-shift and 
phase-shift parameters in waveforms living in different theories. Notice, however,
that we do not minimize over \emph{all} parameters, as would be required to relate
${\cal{N}}_{e}$ to the BF.  A full minimization procedure is costly and it would involve 
an MCMC analysis in general, while the minimization with respect to only $(\Delta t, \Delta \phi)$ is inexpensive. 

We can now see that this quantity is directly related to model selection through the BF: 
\be
\log {\rm BF} \sim 2\pi^{2}{\rm{SNR} }^2 \min_{\lambda^{a}} {\cal{N}}_{e}^{2}\,.
\label{Eq:bf-with-N}
\ee
The effective cycles, $\mathcal{N}_e$, as defined in Eq.~\eqref{Eq:EffCyc}, i.e.~minimized over $(\Delta t, \Delta \phi)$
only, give an \emph{upper limit} to the BF. The fully minimized ${\cal{N}}_{e}$ will in general be smaller than Eq.~\eqref{Eq:EffCyc} due to covariances between system parameters. This means that $\mathcal{N}_e$ is not a perfect proxy for detectability. That is, if $\mathcal{N}_e$ due to a particular term in the phase is large, this may or may not mean that the term is detectable. However, if $\mathcal{N}_e$ due to a particular phase term is small, this \emph{does} indicate that the term will \emph{not} be detectable.

The above is an alternative means of deriving the quantity first derived in \cite{PhysRevD.78.124020}, which is there referred to as the distinguishability/measurability. Up to some rearranging of various factors, this quantity and the effective cycles of phase are the same. The connection between this quantity and the Bayes factor, though, is a new result.

How are the useful and effective cycles related? From the definitions of $\mathcal{N}_u$ and $\mathcal{N}_e$, it is clear that the former gives the difference in the harmonic mean of the number of cycles, while the latter gives the root-mean-square difference in the number of cycles. This is an important difference - $\mathcal{N}_e$ is directly related to the BF in the small deformation limit. For certain signals, however, the difference can be shown to be a GR-modification dependent constant factor. To see this, consider inspiral GWs in the PN approximation both in GR and in ppE form. The orbital phase $\phi(f) = 2 \pi f t(f) - \Psi(f) - \pi/4$ is a power series in $v = (\pi M f)^{1/3}$, just like the Fourier phase $\Psi(f)$ in both GR and ppE theory. The logarithmic derivative, $d\phi/d\ln f$, preserves the structure of such a power series, and so the PN series for $\Psi(f)$ and for $N(f)$ differ only by $b$-dependent, order unity factors in each of the coefficients. The relation between $\mathcal{N}_e$ and $\mathcal{N}_u$ is shown in Fig.~\ref{fig:NuvNe} and derived in Appendix~\ref{AppB}. Of course, the useful and effective cycles defined here can be computed given \emph{any} form of phase evolution, and thus, they are not restricted to phases in the PN approximation.  
\begin{figure}[ht]
\includegraphics[clip=true,angle=-90,width=0.45\textwidth]{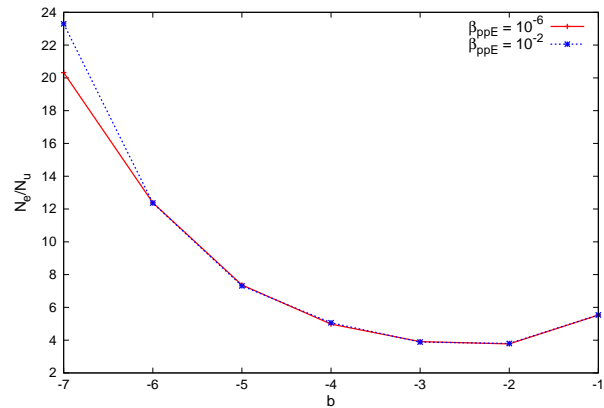} 
\caption{\label{fig:NuvNe} (Color Online) The ratio of effective cycles to useful cycles, calculated as a function of $b$, for two fixed values of $\beta_\ppE$.}
\end{figure}

\subsection{Useful and Effective Cycles as a Measure of Detectability of General non-GR Effects}

Now that we have introduced the concept of effective cycles, we will use it to determine when a particular GR deviation is detectable. In particular, we will determine the number of useful and effective cycles that are needed for ppE deviations to lead to a BF that favors the ppE model.

We inject ppE signals with varying strength $\beta_{\ppE}$ parameters and exponent $b$ parameters and then 
\begin{itemize}
\item[(i)] calculate $\mathcal{N}_u$ and $\mathcal{N}_e$ due to the ppE terms relative to a GR signal ($\beta_{\ppE}=0$) and; 
\item[(ii)] run an MCMC analysis to calculate the BF between a ppE model and the GR model. 
\end{itemize}
The second item requires the choice of a prior range for each ppE strength parameter, which we choose to be $|\beta_\ppE| \le 5\times10^{-5}$ for $b=-7$, $|\beta_\ppE|\le 5\times 10^{-4}$ for $b=-6$, $|\beta_\ppE| \le 5$ for $b=-5$ and $b=-4$. These prior ranges were derived by examining the results in Ref.~\cite{cornish-PPE} and requiring that the deviations be detectable given a GW signal with SNR $\approx 10$. 

Every injection in this study consists of a $(1.4074,1.7415) M_\odot$, NS/NS, non-spinning binary, with zero inclination angle, and with SNR $\approx 15$, which corresponds to a luminosity distance $D_L \approx 50$ Mpc. We choose the zero-detuned, high-power spectral noise density projected for aLIGO~\cite{LIGOnoise}, stopping all integrations at $1000$Hz. This frequency is lower than the GW frequency at which the NSs touch each other, which is approximately between $1250$ and $2000$ Hz, depending on the NS EoS.

\begin{figure*}[ht]
\includegraphics[clip=true,angle=-90,width=0.475\textwidth]{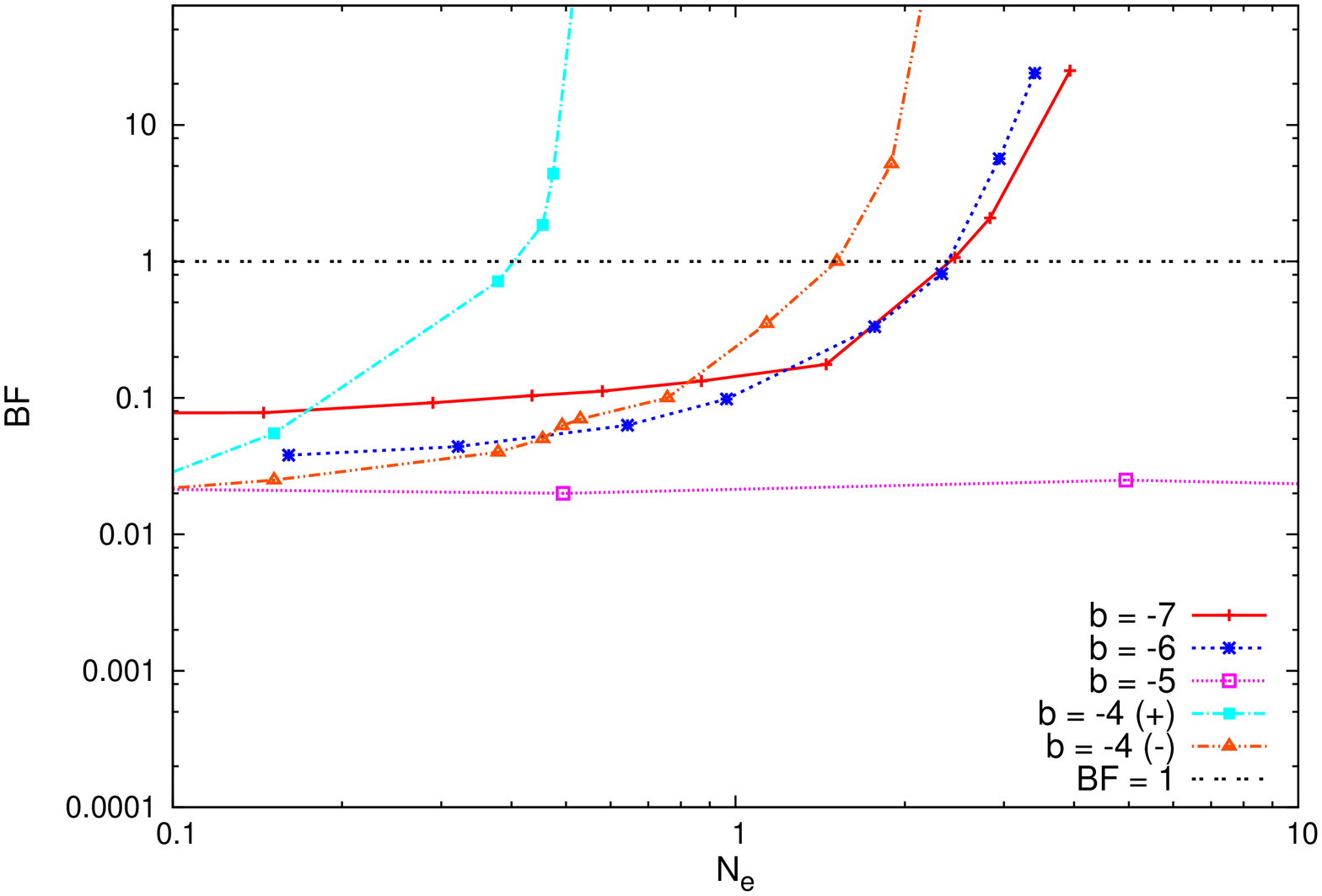} 
\includegraphics[clip=true,angle=-90,width=0.475\textwidth]{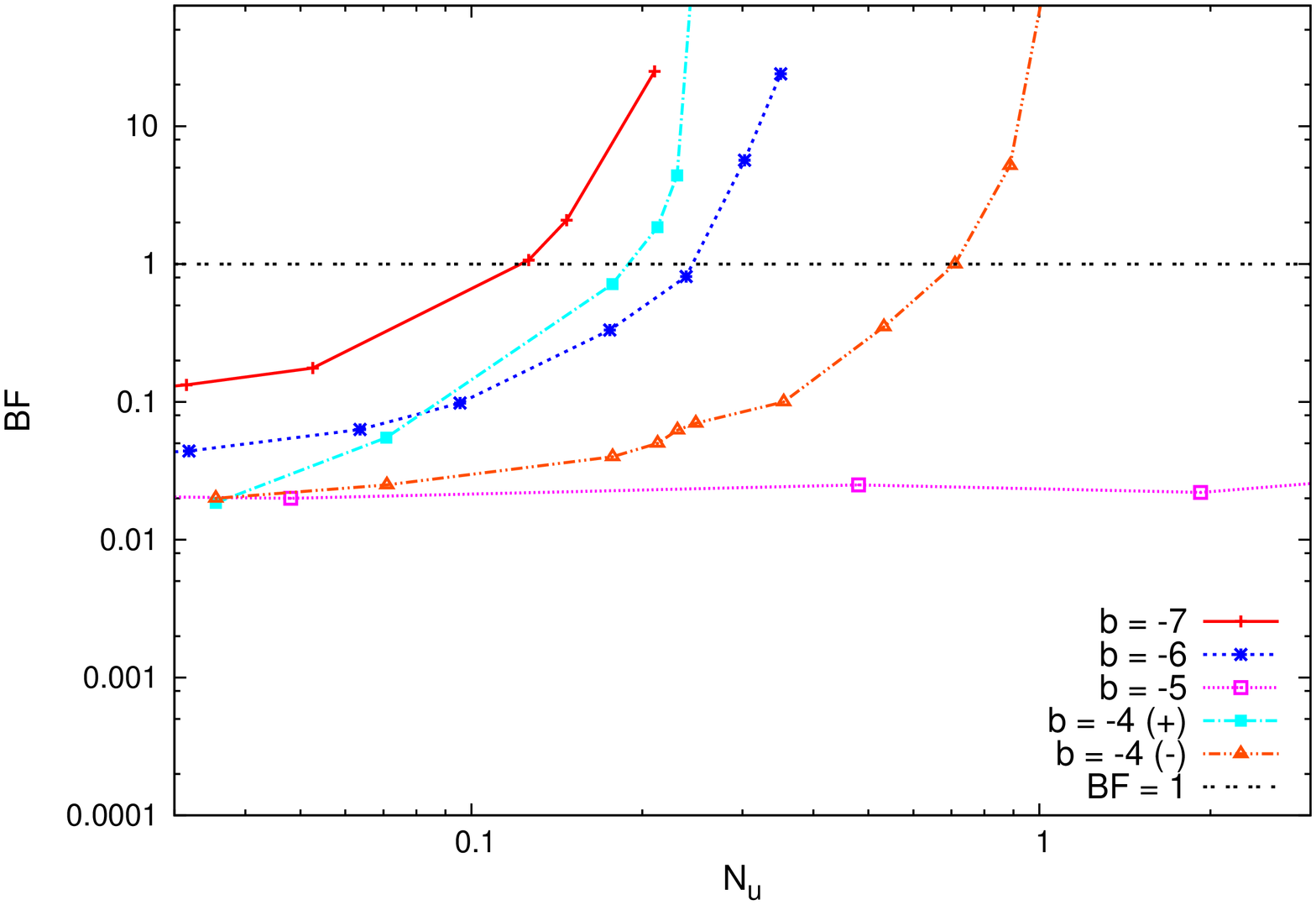} 
\caption{\label{fig:BFNuse} (Color Online) BF in favor of a modification to GR versus the number of effective cycles (left) and useful (right) cycles induced by that modification for different types of ppE corrections. For the $b=-4$ cases, the line labeled (-) corresponds to a negative value for $\beta_\ppE$, and the line labeled (+) corresponds to a positive value; all other injections had positive values for $\beta_\ppE$. Observe that the $b$ dependence of the relationship between BF and $\mathcal{N}_u$ is the opposite of what one would expect.}
\end{figure*}

Figure~\ref{fig:BFNuse} shows the BF on a logarithmic scale versus the absolute value of the number of effective (left panel) and useful (right panel) cycles introduced by signals with different ppE exponent parameters, starting at $b=-7$ (a $-1$PN term) and going up to $b=-4$ (a $0.5$PN term). Recall that ST theories lead to modifications at $-1$PN order and higher and that a BF larger than $1$ indicates the data supports the non-GR model. Observe that detectability occurs when $\mathcal{N}_u$ is between $0.1$ and $1$ and when ${\cal{N}}_{e}$ is between $2$ and $4$ cycles of phase for most of the ppE injections. This is not true, however, for the $b=-5$ case, because of the almost perfect correlation between chirp mass $\mathcal{M}$ and $\beta_\ppE$ when $b=-5$: a straight line in the two-dimensional ${\cal{M}}_{c}$--$\beta_{\ppE}$ plan similar to Fig.~$6$ in~\cite{Sampson:2013lpa}. Because of this, a deviation from GR at the $0$PN (Newtonian) level would have to be very large to be detectable, as previously noted in Ref.~\cite{cornish-PPE}. 

Observe also in Fig.~\ref{fig:BFNuse} the difference in detectability for phase terms that accumulate either positive or negative cycles of phase. As stated, Fig.~\ref{fig:BFNuse} shows the \emph{absolute value} of the useful or the effective cycles of phase introduced by each ppE term. For injections made with a positive $\beta_\ppE$, the actual sign of the cycles of phase is negative. We used this type of injection for most of the lines in this figure, but for the $b = -4$ case, we injected both positive and negative values of $\beta_\ppE$. We find that the positive values are detected more easily than the negative values. 

\begin{figure*}[ht]
\includegraphics[clip=true,angle=-90,width=0.48\textwidth]{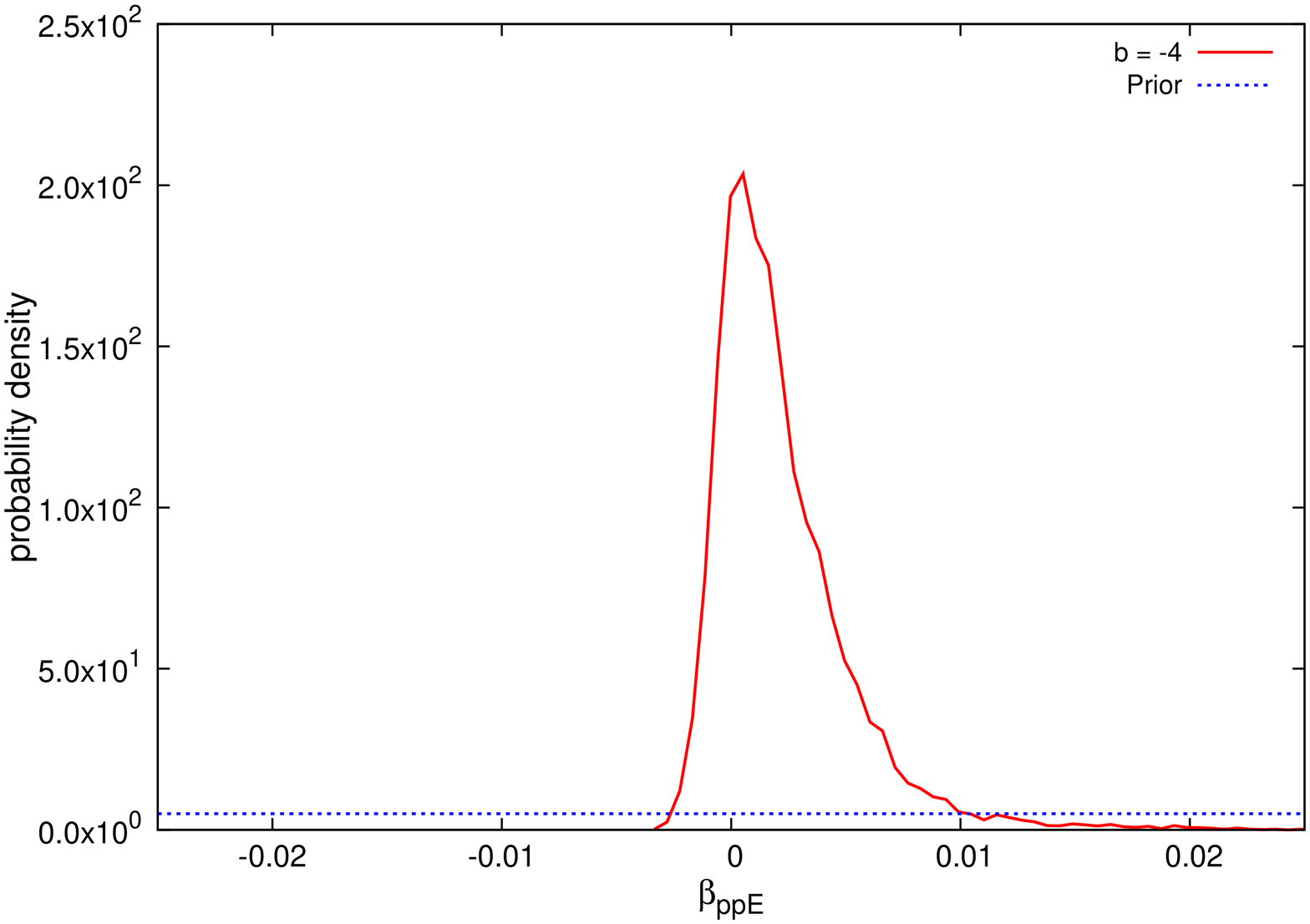} 
\includegraphics[clip=true,angle=-90,width=0.48\textwidth]{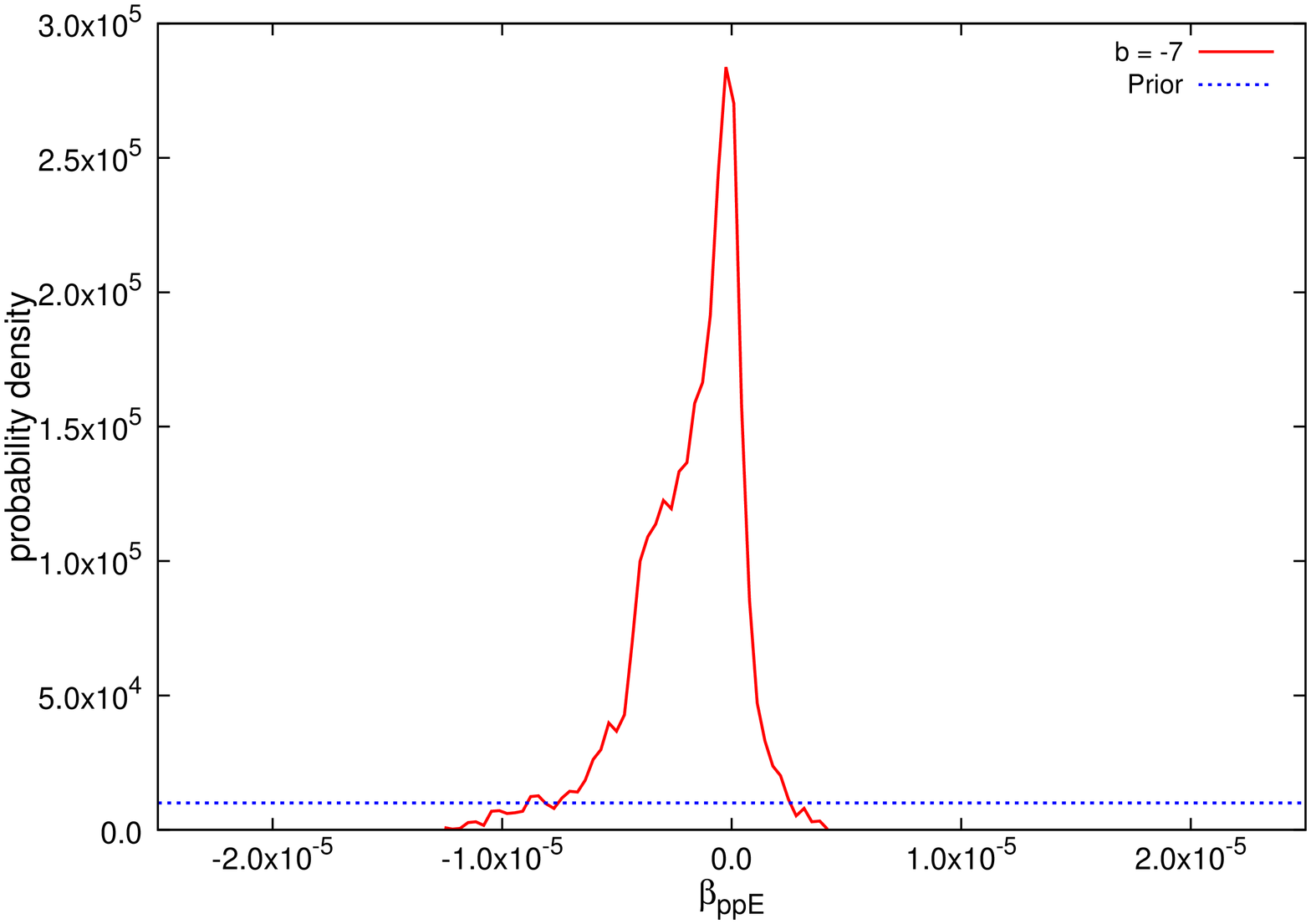} 
\caption{\label{fig:47post} (Color Online) Posterior distributions for $\beta_\ppE$ for a ppE search template with $b=-4$ (left) and $b=-7$ (right), given a GR injection. Observe that these distributions are not symmetric about zero, which explains why changes to the GW phase of different signs are detectable at different levels.}
\end{figure*}

This effect can be understood by examining the posterior distribution of 
$\beta_\ppE$ at $b=-4$, shown in Fig.~\ref{fig:47post} together with the 
posterior of $\beta_\ppE$ at $b=-7$. Observe that the posteriors are not 
symmetric about $\beta_\ppE = 0$. In the case of $b=-4$, injecting a positive 
value of $\beta_\ppE$ results in no posterior weight at $\beta_\ppE = 0$ for 
much smaller values of $|\beta_\ppE|$ than injecting a negative value. The 
asymmetry in the posterior distributions can again be understood by noting that 
$\beta_\ppE$ is correlated with the mass parameters, which are, of course, 
forced to be positive.

Figure~\ref{fig:BFNuse} also shows that the threshold in $\mathcal{N}_{e}$ for detectability (e.g.~the value of $\mathcal{N}_{e}$ at which the BF equals 10) is lower for terms that are of very high PN order. Comparing the left and right panels of this figure, observe that the $\mathcal{N}_{u}$ threshold exhibits the opposite behavior. The $\mathcal{N}_{e}$ threshold behavior is what one would expect because high PN order terms have small covariances with the ppE strength parameters and the system parameters. It is therefore very difficult for a GR waveform to match the phasing of these types of injections, while the opposite is true for low PN order ppE effects. This fact illustrates the advantage of using $\mathcal{N}_e$ instead of $\mathcal{N}_u$ as a measure of detectability: the $\log({\rm{BF}})$ as a function of $\mathcal{N}_e$ exhibits the expected behavior as $b$ changes, but $\log({\rm{BF}})$ as a function of $\mathcal{N}_u$ shows the opposite behavior.

\begin{figure}[ht]
\includegraphics[clip=true,angle=-90,width=0.48\textwidth]{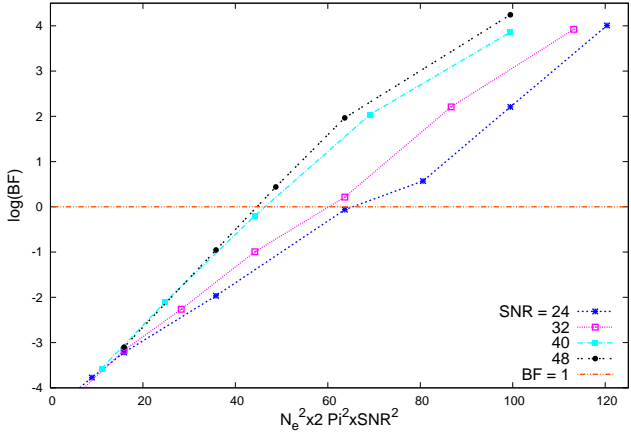} 
\caption{\label{fig:BFvSNRppE} (Color Online) The BF in favor of a deviation from GR as a function of effective cycles of phase scaled by their SNR for ppE injections with $b=-3$. Observe the almost linear relation between log(BF) and $\mathcal{N}_e^2 {\rm{SNR}}^{2}$, as predicted in Eq.~\eqref{Eq:bf-with-N}.}
\end{figure}

So far, the discussion has assumed a fixed SNR of $15$, but clearly the detectability of a GR deviation depends on the SNR of the signal. To understand this, we performed the same analysis as above but with $b=-3$ fixed and signals of differing SNRs. The results are plotted in Fig.~\ref{fig:BFvSNRppE}, where observe that the number of effective cycles necessary for a detection of a modification to GR scales approximately as the SNR squared, as expected from Eq.~\eqref{Eq:bf-with-N}. That equation was derived assuming small deviations from GR, but here we see that regardless of the value of the ppE exponent, Eq.~\eqref{Eq:bf-with-N} is still approximately satisfied.

\subsection{Useful and Effective Cycles as a Measure of Detectability of ST Effects}
\label{sec:useful-effective-ST}

We now apply what we have learned about effective cycles to ST theories. First, 
we must discuss exactly which signals we will analyze in detail. We employ data 
from Ref.~\cite{Palenzuela:2013hsa}, which consists of 80 
GW signals from NS binaries with constituent masses ranging from approximately $1.4$ to 
$1.7M_\odot$, and with $\beta_\ST$ ranging from $-4.5$ to $-3.0$. Within this 
set, and for the polytropic EoS used 
in Ref.~\cite{Barausse:2012da,Palenzuela:2013hsa}, some systems undergo spontaneous/induced scalarization, 
while others dynamically scalarize. As already mentioned, we refer to signals
from systems whose components undergo spontaneous/induced scalarization as ``spontaneously scalarized signals.'' We  do so because a necessary condition for induced scalarization to happen is the presence of at least
one spontaneously scalarized star. The term ``dynamically scalarized signals'' will
denote those from dynamically scalarized binaries.
The strength and type of scalarization that occurs depends sensitively on the constituent masses and the 
value of $\beta_{\ST}$.

We now wish to calculate the effective cycles $\mathcal{N}_e$ induced by the ST 
corrections to investigate their detectability. For the spontaneously scalarized 
cases, we can compute $\mathcal{N}_e$ with the SPA waveforms, using for $\Delta 
\Psi(f)$ the terms in square brackets of Eq.~\eqref{eq:SPAphase}. The integrated SPA
dephasing is then minimized over $(\Delta t,\Delta \phi)$, as explained in Eq.~\eqref{Eq:EffCyc}.
For the dynamically scalarized cases, we compute $\mathcal{N}_e$ with the Fourier transform 
of the numerical time-domain data of Ref.~\cite{Palenzuela:2013hsa}, 
using for $\Delta \Psi(f)$ the difference in the Fourier phases of a GR and a ST numerical signal.
We then again introduce parameters $(\Delta t,\Delta \phi)$ and minimize the integrated dephasing with
respect to them to define ${\cal{N}}_{e}$, as explained in Eq.~\eqref{Eq:EffCyc}. 

\begin{figure*}[ht]
\includegraphics[clip=false,angle=-90,width=0.48\textwidth]{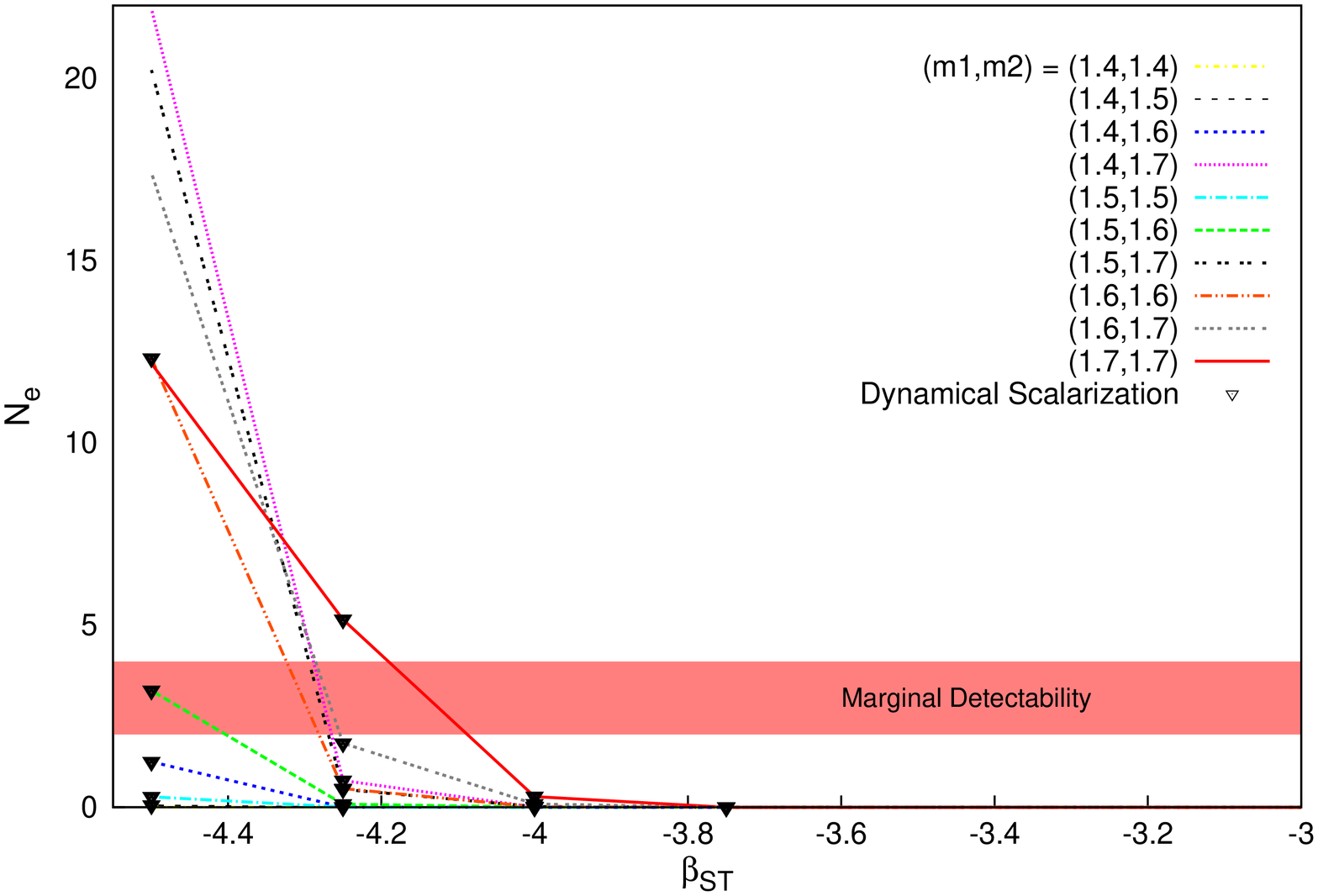}
\includegraphics[clip=false,angle=-90,width=0.48\textwidth]{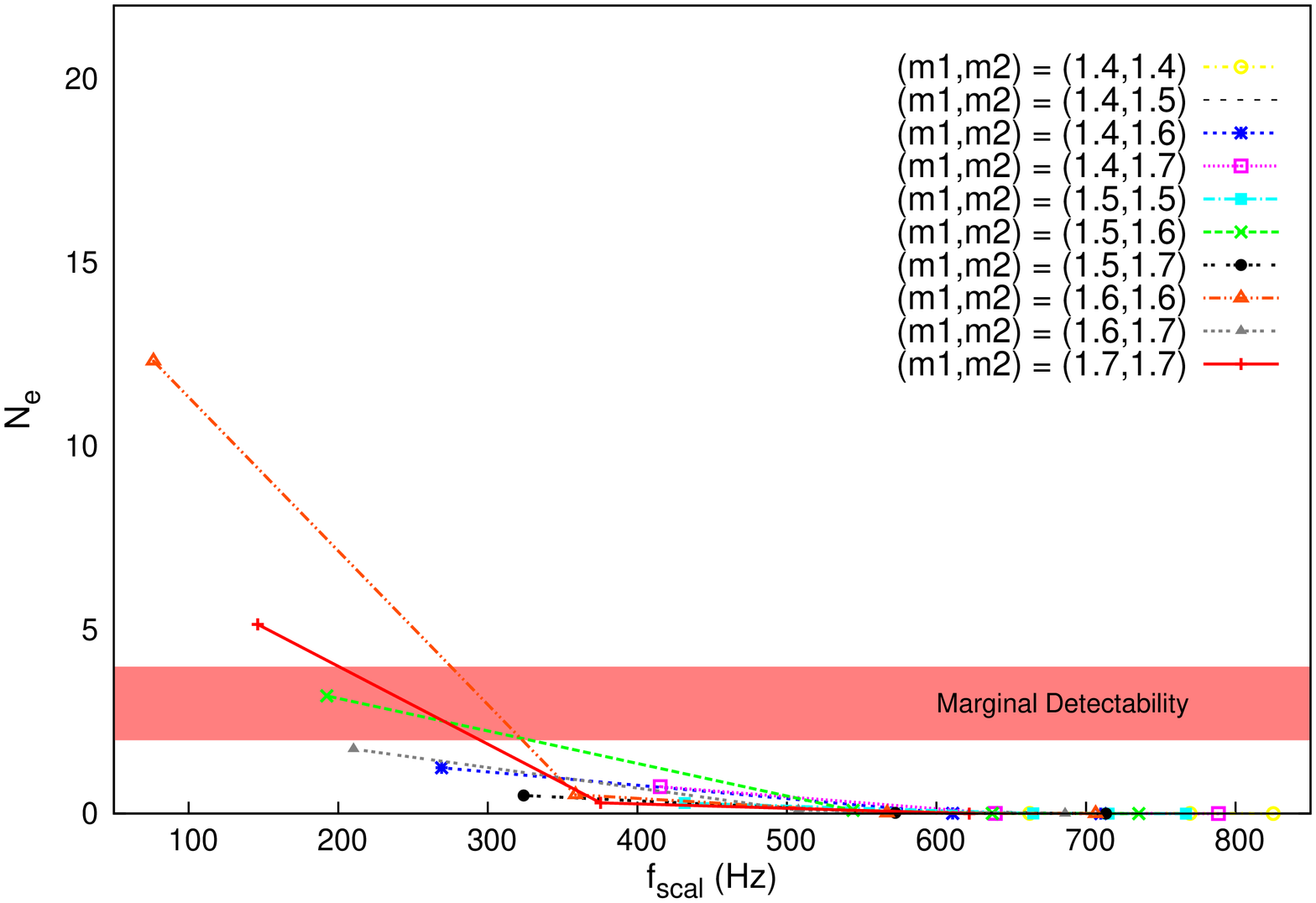}
\caption{\label{fig:BFvbetaST} (Color Online) Effective number of cycles generated by a ST modification to GW signals as a function of $\beta_\ST$ for all ST cases (left) and as a function of GW frequency of scalarization in the dynamical scalarization cases (right). Different curves correspond to systems with different masses, and the different points in the right panel correspond to different values of $\beta_{\ST}$ increasing to the right. We model the GW phase through the SPA in the spontaneously scalarized cases, while we use the Fourier transform of numerical data in the dynamically scalarized cases. The shaded region corresponds roughly to the number of effective cycles necessary for BFs between $1$ and $10$, as estimated from Fig.~\ref{fig:BFNuse}. Observe that the number of effective cycles is below the detectability region for all but the $\beta_\ST \lesssim -4.5$ cases. Note also that the only dynamically scalarized cases that are detectable are those in which the scalar field activates at small frequencies. For the equal-mass binaries that undergo spontaneous scalarization, the SPA is not applicable for calculating $\mathcal{N}_e$. We therefore used masses that were nearly, but not identically, equal for an approximate calculation.}
\end{figure*}
The left panel of Fig.~\ref{fig:BFvbetaST} shows the effective cycles as a function of $\beta_{\ST}$ for both spontaneously and dynamically scalarized cases, with the latter labeled with an upside-down triangle. Different line styles correspond to systems with different masses. The shaded region corresponds to the region where one would expect BFs of between $1$ and $10$, given the results of Fig.~\ref{fig:BFNuse}. Modifications that lead to effective cycles above this shaded region may then be detectable with an aLIGO instrument.

Several features of this figure are worth discussing in more detail. 
First, observe that almost all of the detectable cases correspond to spontaneously scalarized systems. For these systems, dipole radiation is the dominant GR modification, a $-1$PN order effect that is proportional to the \emph{difference} of scalar charges of the two bodies (see e.g.~Eq.~\eqref{eq:SPAphase}). For equal-mass binaries, this dipolar effect vanishes identically, and the dominant modification enters at Newtonian, $0$PN order. Such a GR modification, however, is strongly degenerate with the chirp mass, as shown in Fig.~\ref{fig:BFNuse}, and thus, it is difficult to detect.  For this reason, spontaneously scalarized systems with larger mass differences lead to larger values of $\mathcal{N}_e$ and are easier to detect.

\begin{figure}[ht]
\includegraphics[clip=false,angle=-90,width=0.48\textwidth]{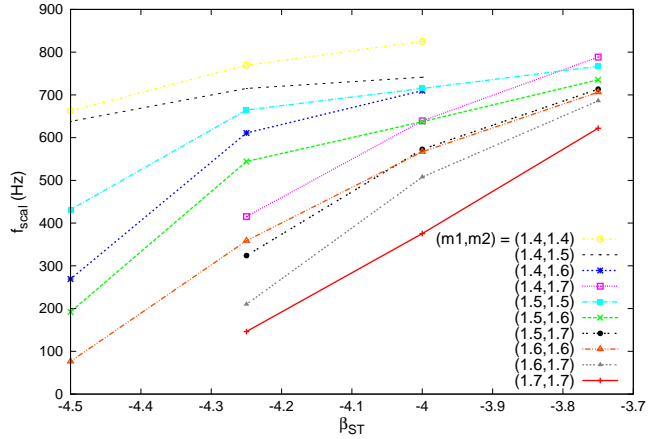}
\caption{\label{fig:fstarvbeta} (Color Online) Approximate GW frequency in Hz at which the scalar field activates in a dynamically scalarized binary. Observe that the higher $\beta_{\ST}$ and the lower the total mass of the system, the higher the frequency of activation }
\end{figure}

Another interesting feature of Fig.~\ref{fig:BFvbetaST} is that, within the cases analyzed, only a handful of dynamically scalarized systems seem detectable: the $(m_{1},m_{2}) = (1.6,1.6) M_{\odot}$ with $\beta_{\ST} = -4.5$ system, the $(m_{1},m_{2}) = (1.7,1.7) M_{\odot}$ with $\beta_{\ST} = -4.25$ system and the $(m_{1},m_{2}) = (1.5,1.6) M_{\odot}$ with  $\beta_{\ST} = -4.5$ systems. One of the key differences between these cases and all others is that they scalarize at relatively low GW frequency. Figure~\ref{fig:fstarvbeta} shows the approximate GW frequency at which the scalar field 
activates in a dynamically scalarized binary, as a function of $\beta_{\ST}$. Observe that as $|\beta_{\ST}|$ becomes smaller, or as the total mass of the binary decreases, dynamical scalarization occurs at higher and higher frequencies. For these three cases the scalar field activates at roughly $80$, $120$, and $180$ Hz respectively, while in all other cases dynamical scalarization occurs at higher GW frequency.

The reason, then, that the three dynamically scalarized cases discussed above appear detectable is that detectability of a \emph{sudden} non-GR effect, i.e.~one that turns on rapidly, correlates strongly with the GW frequency at which this turn on occurs. The right panel of Fig.~\ref{fig:BFvbetaST} shows this correlation through $\mathcal{N}_{e}$ as a function of the GW frequency of dynamical scalarization. Notice that the lower the GW frequency of activation, the larger the number of effective cycles, and thus, the easier it would be to detect such a GR modification.

Reference~\cite{Sampson:2013jpa} first observed this phenomenon by studying ppE-type GR modifications that turn on suddenly. Their conclusion was that for aLIGO to detect such non-GR effects at SNRs of $12$, the modification had to turn on at a GW frequency lower than roughly 100 Hz (see Fig.~$4$ in~\cite{Sampson:2013jpa}). The ST modifications we study here are of a different PN order and at higher SNR than the non-GR ppE signals considered in Ref.~\cite{Sampson:2013jpa}, but we still see similar behavior: dynamical scalarization is only detectable when it activates below $\approx 200$ Hz at SNR 15. 

Summing up, if one detected a GW signal that is consistent with GR (i.e.~lacking any dynamical, induced or
spontaneous scalarization effects), one should be able to constrain $\beta_{\ST} \lesssim -4.25$, 
given the data that we studied in this paper.
This statement, of course, is EoS dependent, and thus, 
strictly applicable only to NSs with the polytropic EoS used here. In principle, NS binaries 
with a different EoS could scalarize
at a different GW frequency. With some variability depending on the EoS, though, it remains true that one must have $\beta_\ST \lesssim -4.25$ in order for scalarization
(either spontaneous/induced or dynamical) to occur at all.

\begin{table*}[!]
\begin{tabular}{c||c|c|l}
        \hline
	Case	&       Masses  (in $M_\odot$)	&	Theory	&	Type of Scalarization \& Radiation	\\
        \hline\hline
	a	& $(1.4074,1.4074)$ &       $\tilde{\beta}=(-4.5, -4.2$)
		& Dynamical Scalarization	\\
		&	&	&	Quadrupolar, No Dipolar Radiation
	\\
        \hline
	b	& $(1.4074,1.5145)$ &       $\tilde{\beta}=(-4.5, -4.2$)
		& Dynamical Scalarization	\\
		&	&	&	Quadrupolar \& Dipolar Radiation
	\\
        \hline
	c	& $(1.4074,1.6441)$ &       $\tilde{\beta}=(-4.5, -4.2$)
		& Dynamical/Induced Scalarization	\\ 
		&	&	&	Quadrupolar \& Dipolar Radiation
	\\
        \hline
	d	& $(1.4074,1.7415)$ &       $\tilde{\beta}=(-4.5, -4.2$)
		& Dynamical/Induced Sc. for the lower-mass star	\\
		&	&	& Spontaneous Scalarization in the higher-mass star	\\
		&	&	&	Quadrupolar \& Dipolar Radiation
	\\
        \hline
	e	& $(1.5145,1.5145)$ &       $\tilde{\beta}=(-4.5, -4.2$)
		& Dynamical Scalarization	\\
		&	&	&	Quadrupolar, No Dipolar Radiation
	\\
        \hline
	f	& $(1.5145,1.6441)$ &       $\tilde{\beta}=(-4.5, -4.2$)
		& Dynamical Scalarization	\\
		&	&	&	Quadrupolar \& Dipolar Radiation
 	\\ 
        \hline
	g	& $(1.5145,1.7415)$ &       $\tilde{\beta}=(-4.5, -4.2$)
		& Dynamical/Induced Sc. for the lower-mass star	\\
		&	&	& Spontaneous Scalarization in the higher-mass star	\\
		&	&	&	Quadrupolar \& Dipolar Radiation
	\\
        \hline
	h	& $(1.6441,1.6441)$ &       $\tilde{\beta}=(-4.5, -4.2$)
		& Dynamical/Induced Scalarization		\\
		&	&	&	Quadrupolar, No Dipolar Radiation
 	\\ 
        \hline
	i	& $(1.6441,1.7415)$ &       $\tilde{\beta}=(-4.5, -4.2$)
		& Spontaneous Scalarization for $\tilde{\beta} = -4.5$	\\
		&	&	&	Dynamical/Induced Scalarization for $\tilde{\beta} = -4.2$	\\
		&	&	&	Quadrupolar \& Dipolar Radiation
	\\
	\hline
	j	& $(1.7415,1.7415)$ &       $\tilde{\beta}=(-4.5, -4.2$)
		& Spontaneous Scalarization for $\tilde{\beta} = -4.5$	\\
		&	&	&	Dynamical/Induced Scalarization for $\tilde{\beta} = -4.2$	\\
		&	&	&	Quadrupolar, No Dipolar Radiation
	\\
        \hline\hline
\end{tabular}
\caption{System parameters for the $\beta_\ST = -4.5$ and $\beta_\ST = -4.25$ cases discussed in this section.}
\end{table*}

\section{Detectability of Scalar-Tensor Deviations: Bayesian Model Selection}
\label{sec:BF}

In this section, we carry out the second part of our data analysis investigation to determine the detectability of ST deviations in GWs emitted during the inspiral of NS binaries. We perform a full Bayesian analysis study, separating the spontaneously scalarized cases from the dynamically scalarized ones. Such a study will allow us to confirm the expectations derived using effective cycles in the previous section. 

To test these expectations, we inject ST GW signals at an SNR $\approx 15$ produced by
\begin{itemize}
\item[(a)] {\bf{Spontaneously scalarized NS binaries}}, with spontaneous/induced scalarization occurring \emph{before} GWs enter the detector's sensitivity band;
\item[(b)] {\bf{Dynamically scalarized NS binaries}}, with dynamical scalarization occurring \emph{during} the inspiral, at GW frequencies in the detector's sensitivity band. 
\end{itemize}
As a case study for spontaneously scalarized injections, we consider a $(1.4074,1.7415) M_{\odot}$ binary with $\beta_{\ST} \ge -4.25$. We expect these signals to lead to the largest spontaneous scalarization effects, as one can see in the left panel of Fig.~\ref{fig:BFvbetaST} at $\beta_{\ST} = -4.5$. However, for $\beta_{\ST} > -4.25$, we do not expect these effects to be detectable, since they lead to a very small number of effective cycles. Note that for $\beta_{\ST} \to -4.25$, this system is a case that exhibits dynamical scalarization. 

For dynamically scalarized injections, we consider several different systems. First, we study a $(1.4074,1.7415) M_{\odot}$ binary at $\beta_{\ST} = -4.25$, since this is the limiting case of the spontaneously scalarized sequence discussed in the previous paragraph. We then study a $(1.6441,1.6441) M_{\odot}$ binary at $\beta_{\ST} = -4.5$, a $(1.7415,1.7415) M_{\odot}$ binary at $\beta_{\ST} = -4.25$, and a $(1.5145,1.6441) M_{\odot}$ binary at $\beta_{\ST} = -4.5$, as these are the dynamically scalarized signals that look the most detectable, given the left panel of Fig.~\ref{fig:BFvbetaST}. 

Spontaneously and dynamically scalarized injections are modeled differently. For the former, we use the SPA scheme of Eq.~\eqref{eq:SPAphase}. For the latter, we first compute the discrete Fourier transform of the numerical data of Ref.~\cite{Palenzuela:2013hsa}. We then take the difference of the Fourier phase between a numerical ST signal and a GR signal, and finally add this phase difference to a GR SPA signal.  

We recover these injections with four different types of template families: 
\begin{enumerate}
\item[(i)] {\bf{Simple ppE templates}}, constructed with a single $\beta_{\ppE}$ parameter and ppE exponent $b=-7$; 
\item[(ii)] {\bf{ST SPA templates}}, constructed from the results presented in Sec.~\ref{sec:SPAtemp};
\item[(iii)] {\bf{2-parameter ppE templates}}, an augmented ppE template family that uses two ppE terms, one with exponent $b=-7$ and one with $b=-6$~\cite{Sampson:2013lpa}; 
\item[(iv)] {\bf{ppE$_\theta$ templates}}, another augmented ppE template family with a single ppE exponent $b = -7$ but with $\beta_{\ST} \to \Theta(f-f^*) \beta_\ppE$, where $\Theta(\cdot)$ is a step-function and the threshold frequency, $f^*$, is a new ppE parameter~\cite{Sampson:2013lpa}. 
\end{enumerate}
The simple ppE template family fixes the ppE exponent to $b=-7$, as this corresponds to the leading-order ST correction to the SPA phase for unequal mass systems where dipolar radiation is present. The ST SPA templates, of course, are the same templates as the model used for the spontaneously scalarized injections, and thus, by construction, we expect these to be the best templates for extracting ST modifications of this type. The 2-parameter ppE template family is also able to achieve a perfect match with spontaneously scalarized injected signals, but it includes two free parameters, rather than one. The ppE$_{\theta}$ template family allows the non-GR terms in the phase to ``turn-on'' at a particular threshold frequency, which is well-suited to dynamically scalarized injections. 

A salient feature of the semi-analytical results of Ref.~\cite{Palenzuela:2013hsa} (and of the full general-relativistic simulations of Ref.~\cite{Barausse:2012da}) is the binary's early plunge due to the activation of dynamical scalarization. Such a feature is present in the dynamically scalarized injections we consider, but given the limited number of data sets, we cannot study its detectability in sufficient detail. In order to study whether such rapid termination of the inspiral is detectable, we will consider an additional type of injection and template: 
\begin{enumerate}
\item[(v)] {\bf{Heaviside signal}} of the form $\tilde{h}_{\GR}(f) \Theta(f^{*} - 
f)$\,,
\end{enumerate} 
where we will vary $f^{*}$ within $(40,10^{3})$ Hz. Given such injections and templates, we then study the range of $f^{*}$ that leads to early terminations that can be detected as a non-GR effect. We will explain in Sec.~\ref{dyn-sca-subsec} why we choose to work with such a toy-model. 

All the template models considered above are nested. This means that for a 
certain choice of non-GR parameters, the templates reduce exactly to GR. For 
example, when $\beta_{\ST} = 0$, the ST SPA templates reduce exactly to the GR 
SPA templates. When this is the case, one can use the Savage-Dickey density 
ratio to calculate the BF~\cite{dickey1971weighted}. Finding this ratio requires 
the calculation of the posterior at the nested value of the non-GR parameter. 
The posterior is calculated with MCMC techniques well-developed in previous 
studies~\cite{TysonThesis,cornish-PPE,Sampson:2013jpa,Sampson:2013lpa}. We again 
use the zero-detuned, high-power aLIGO noise curve, assuming a single detector 
and truncating all integrals at $f = 1000$ Hz.

\subsection{Spontaneously Scalarized Signals}

\begin{figure}[ht]
\includegraphics[clip=true,angle=-90,width=0.45\textwidth]{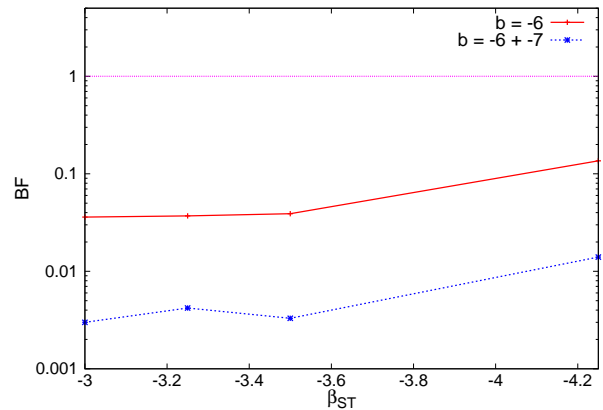} 
\caption{\label{fig:BFSPA} (Color Online) The BF in favor of a non-GR signal, calculated by injecting ST signals with SNR of 15, and recovering using both simple ppE templates, and 2-parameter ppE templates. A BF above 1 indicates the data prefers the non-GR model. As expected, neither template is able to detect the non-GR deviations.}
\end{figure}

We first consider spontaneously scalarized signals and compute the BFs between GR and the first two types of templates described above [(i) and (ii)]. Figure~\ref{fig:BFSPA} shows the BFs as a function of the injected $\beta_\ST$, keeping $(m_{1},m_{2})=(1.4074,1.7415) M_{\odot}$ fixed and varying $\beta_{\ST}$ with the constraint $\beta_{\ST} \ge -4.25$, so as to consider only spontaneously scalarized signals.  Because there are no BFs larger than one, this figure shows that spontaneously scalarized ST deviations from GR are not detectable for any of the cases shown. This is consistent with our expectations from the previous subsection. The BFs for the 2-parameter ppE templates are lower than those shown in Fig.~\ref{fig:BFSPA} for the simple ppE templates because of the Occam penalty for more complicated models, which we will discuss later on in this subsection.

\begin{figure*}[ht]
\includegraphics[clip=true,angle=-90,width=0.45\textwidth]{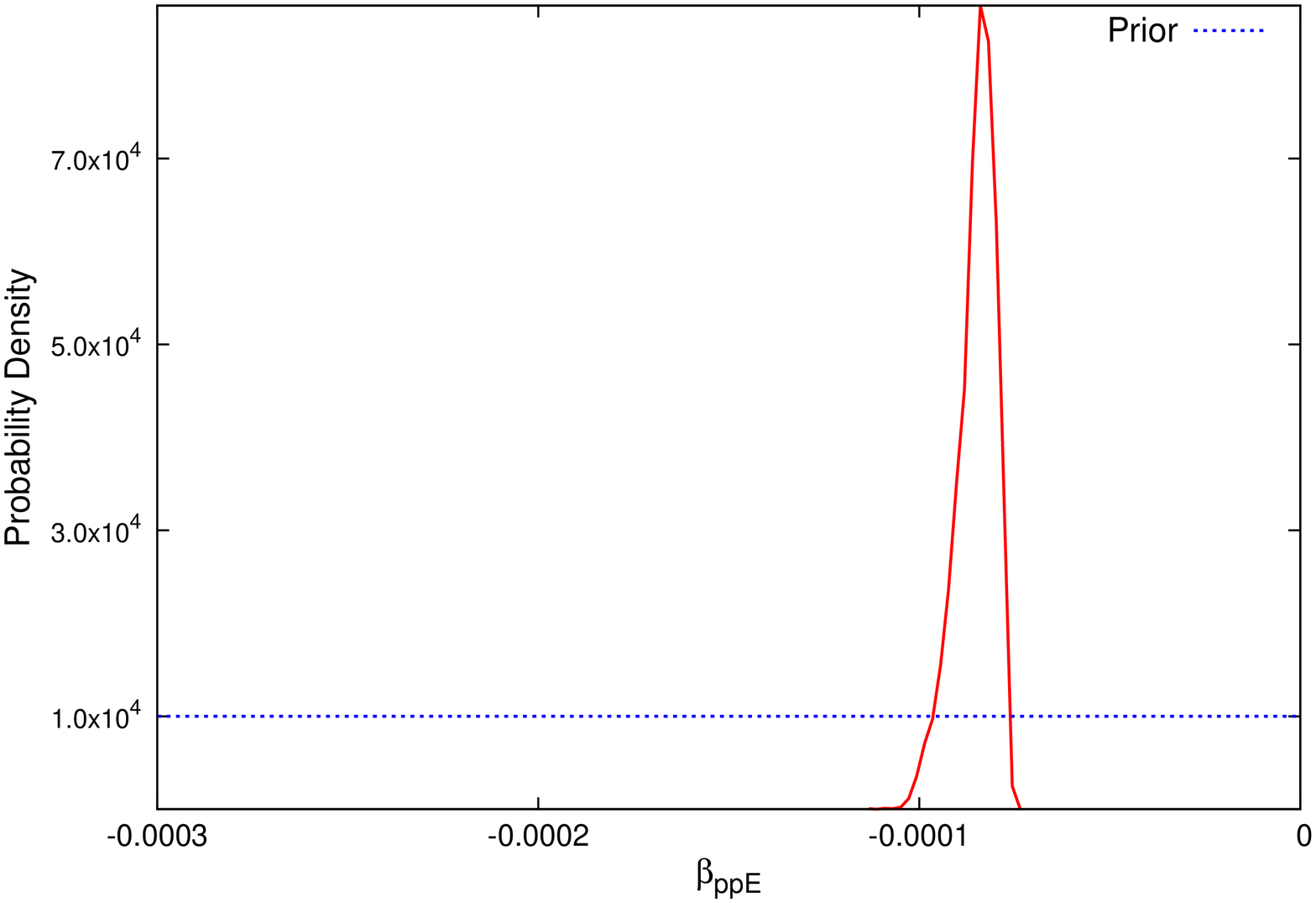}
\includegraphics[clip=true,angle=-90,width=0.45\textwidth]{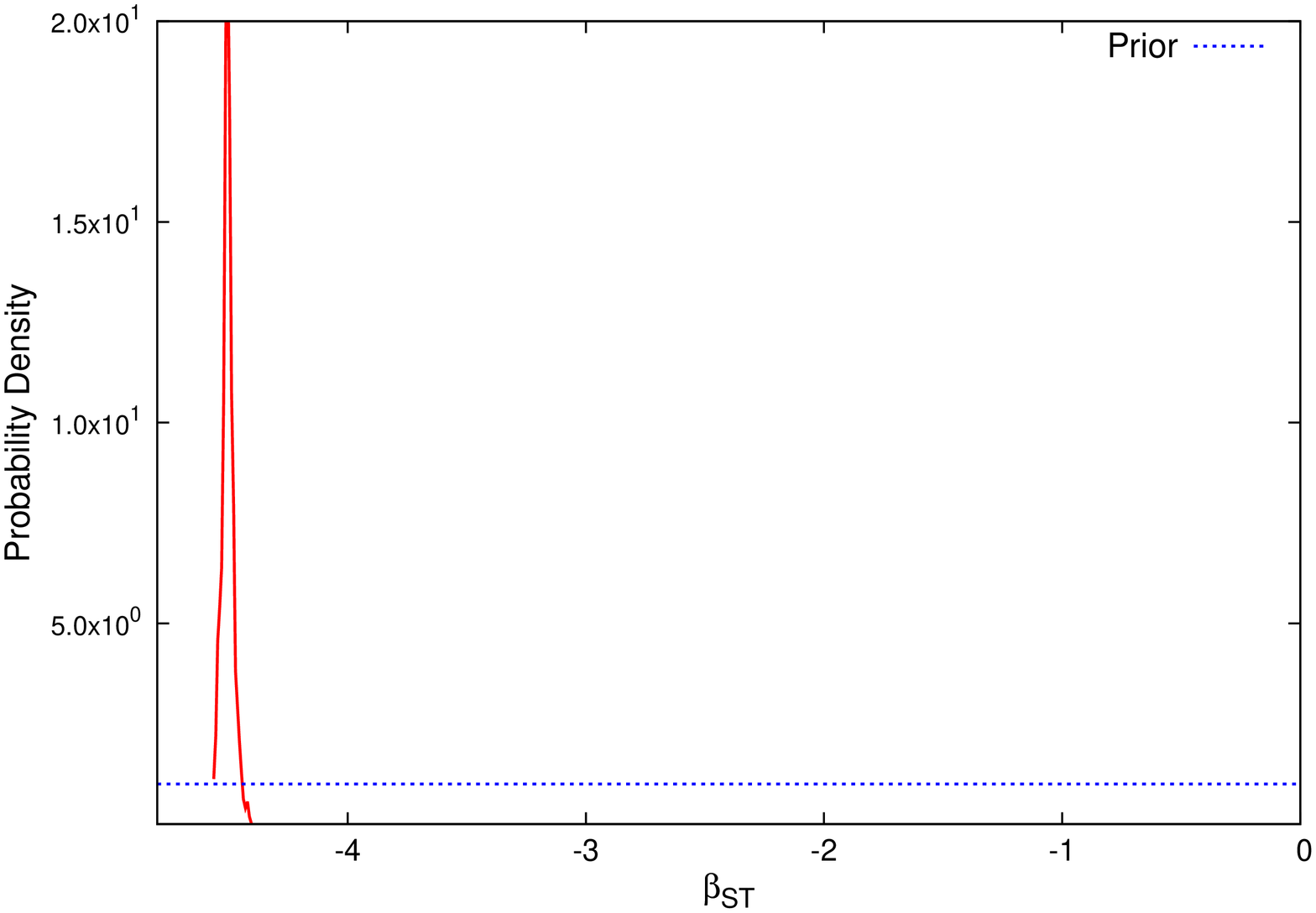}
\caption{\label{fig:beta45} (Color Online) Red (solid) lines: the posterior distributions for $\beta_\ppE$ (left panel) and $\beta_\ST$ (right panel). Blue (dashed) lines: the prior density for each of these parameters. There is essentially zero posterior weight at $\beta_{\ST/\ppE} =0$, indicating a large preference for the non-GR model from both template families.}
\end{figure*}

The BF for the spontaneously scalarized, $\beta_\ST = -4.5$ case is not included in Fig.~\ref{fig:BFSPA} because there is so little posterior weight at $\beta_\ppE = 0$ that a calculation of the BF using the Savage-Dickey density ratio is poorly defined. That is, the Savage-Dickey density essentially diverges due to poor exploration of the $\beta_{\ppE} = 0$ region. Figure~\ref{fig:beta45} shows the posterior distribution for $\beta_\ppE$, which illustrates this point and, as expected, indicates that the spontaneously scalarized binary with $\beta_\ST = -4.5$ is easily detectable, for the polytropic EoS models considered here. 

\begin{figure*}[ht]
\includegraphics[clip=true,angle=-90,width=0.45\textwidth]{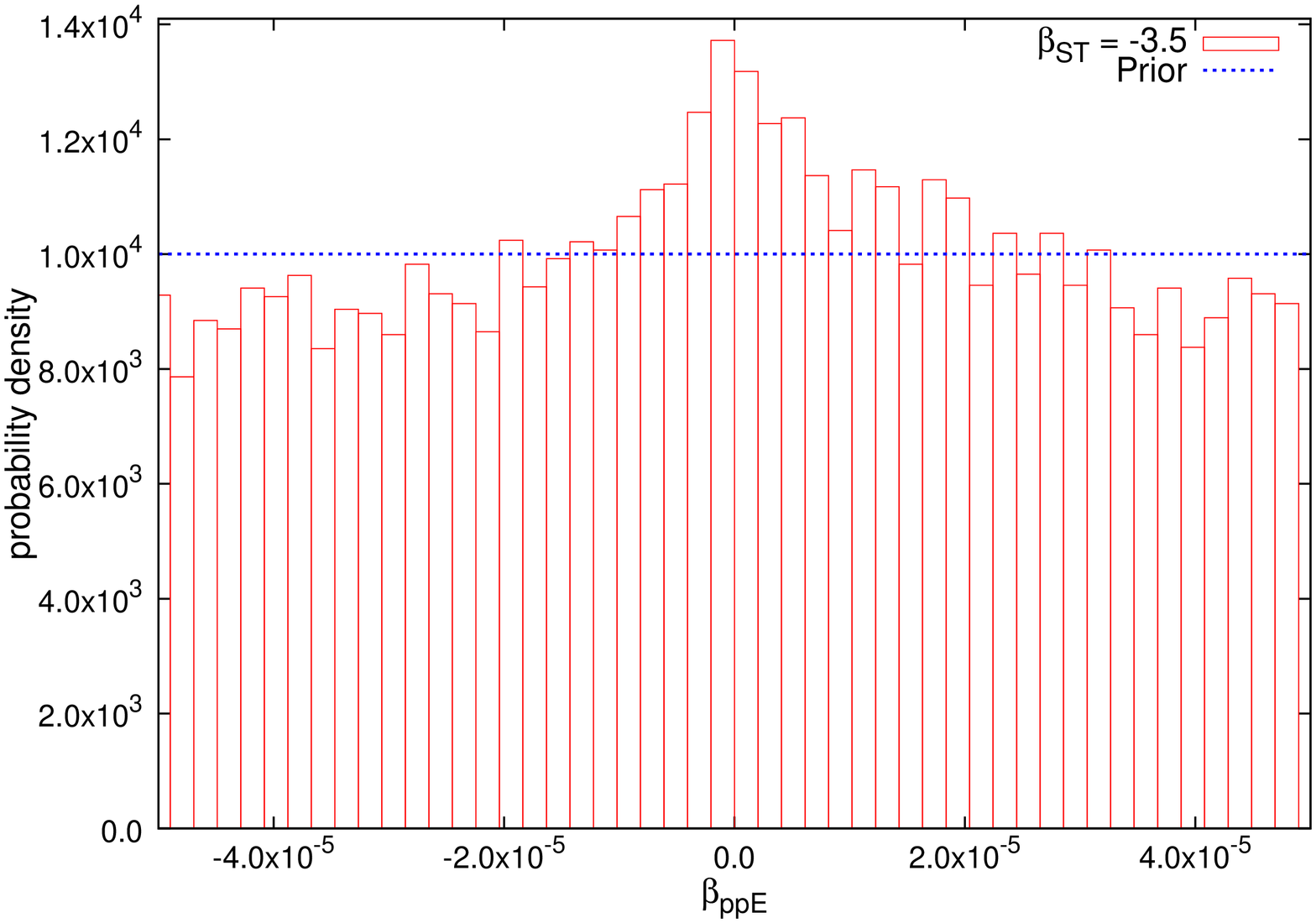} 
\includegraphics[clip=true,angle=-90,width=0.45\textwidth]{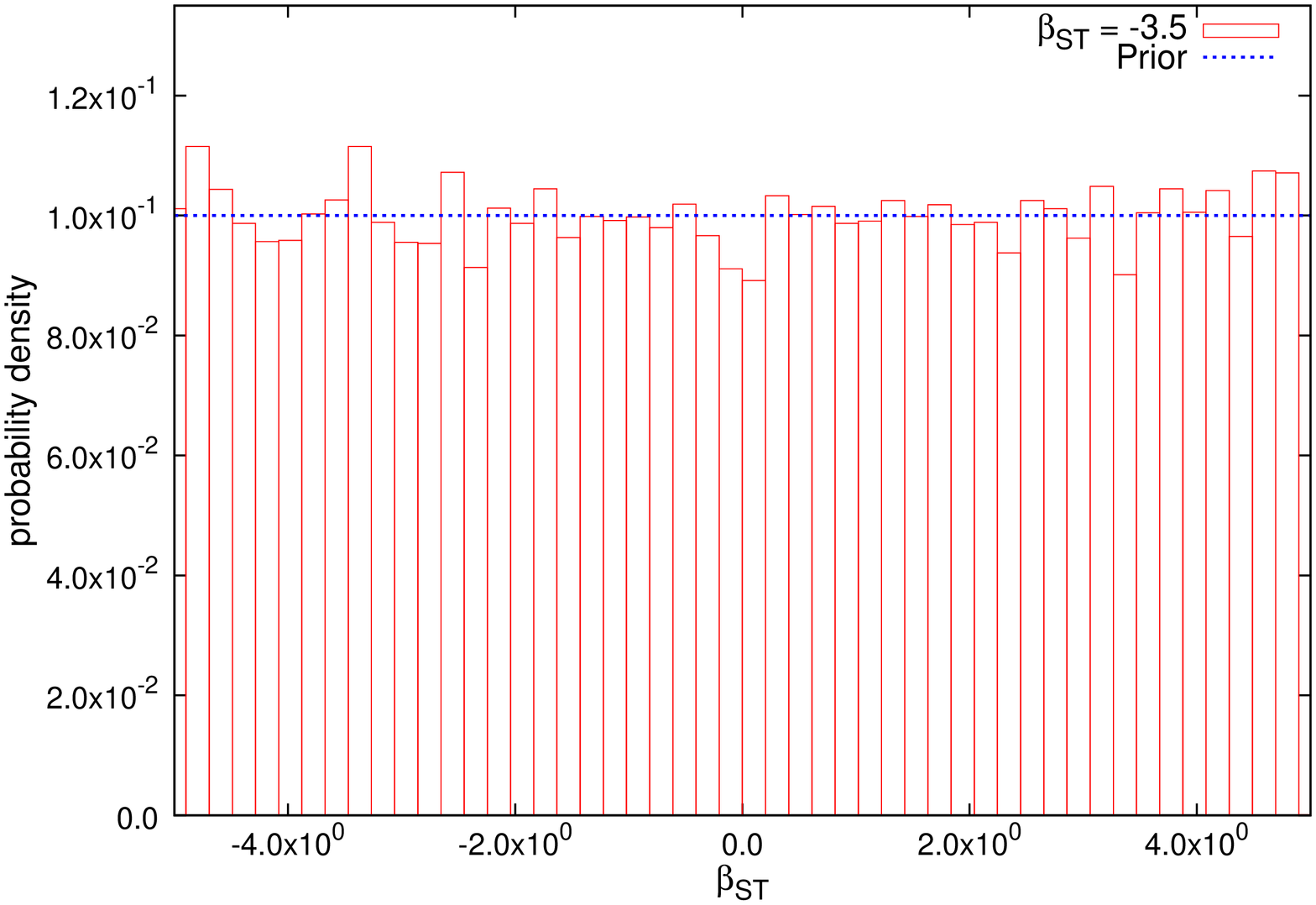} 
\caption{\label{fig:betathetapost} (Color Online) Posterior distributions for $\beta_\ST$ (right) and for $\beta_\ppE$ (left), recovered from an SNR 15 injection with $\beta_\ST = -3.5$. Also plotted in blue (dashed) is the prior density for both parameters. The parameters are essentially unconstrained within their prior range, leading to a BF $\approx 1$ for both models, indicating no preference between this model and GR.}
\end{figure*}

One may worry that our inability to detect spontaneously scalarized binaries when $\beta \geq -4.25$ is somehow a consequence of using ppE templates. To prove that this is not the case, we next explore the extraction of such spontaneously scalarized signals using both ppE$_\theta$ and custom-made, SPA templates. The ppE$_\theta$ templates should have more freedom to fit these signals, and the SPA templates can fit them perfectly. Because of the nature of these waveforms and the weakness of the GR deviation in the injections, the non-GR parameters in both cases are essentially unconstrained within their prior ranges. For the SPA templates, the non-GR parameter is $\beta_\ST$, and its prior range is $|\beta_\ST| < 5$. This means that the BF is approximately equal to one in both cases, independent of the injected parameters. Figure~\ref{fig:betathetapost} shows the prior and posterior distributions generated using the ppE$_\theta$ and the SPA templates, for a spontaneously scalarized binary with $\beta_\ST = -3.5$. These distributions are identical for other injected values of $\beta_\ST$, barring $\beta_\ST \leq -4.5$, which is again easily extractable as a non-GR signal with either type of template. 

We expect, though, that custom-made templates should perform better than ppE templates at extracting non-GR modifications when the custom-made templates match the signal. The extent to which this is true depends on the strength of the non-GR modification and on the loudness of the signal. Let us first consider very strong ST modifications, which we have already shown to be detectable with simple ppE templates, i.e.~a $\beta_\ST = -4.5$, spontaneously scalarized ST signal, with masses of $(1.4074,1.7415) M_\odot$ and SNRs of 12, 10, 8, and 6. When the SNR is 12, 10, or 8, there is \emph{no} difference in the ability of simple ppE or SPA templates to discern the presence of a ST effect. When the SNR drops below 8, the signal is not detectable in the first place, using either type of template. Thus, custom-made templates and model-independent templates are equally good at detecting this type of GR deviations.

But what about GR deviations that are weaker, and thus, more difficult to 
detect? Surely, in this case one expects custom-made templates to be more 
effective at discerning such deviations. To explore this question, we inject a 
$\beta_{\ST}=-3.5$, spontaneously scalarized ST signal, with masses 
$(1.4074,1.7415) M_\odot$ and \emph{very} high SNRs (so that the non-GR 
modifications are detectable). We then recover these signals using both the SPA 
templates and the simple ppE templates. For the former, we again use the prior 
range on $\beta_\ST$ of $|\beta_\ST| \le 5$, while for the latter the prior 
range on $\beta_{\ppE}$ is a bit tricker. We could use the same prior range on 
$\beta_{\ppE}$ as in the previous subsection, but this was motivated from a 
study of signals at ${\rm{SNR}} \approx 20$. The bounds on $\beta_\ppE$ for the 
extremely high SNR signals we are studying in this subsection should be much 
stronger. We estimate the latter by relating the $-1$PN coefficient from our SPA 
waveform to the ppE strength parameter when $b=-7$. This leads to a prior range 
on $\beta_\ppE$ of $|\beta_\ppE| \le 1.035\times 
10^{-10}$. We then calculate the BF for the simple ppE 
model using both the full and the more restricted prior range on $\beta_{\ppE}$.

\begin{figure}[ht]
\includegraphics[clip=true,angle=-90,width=0.45\textwidth]{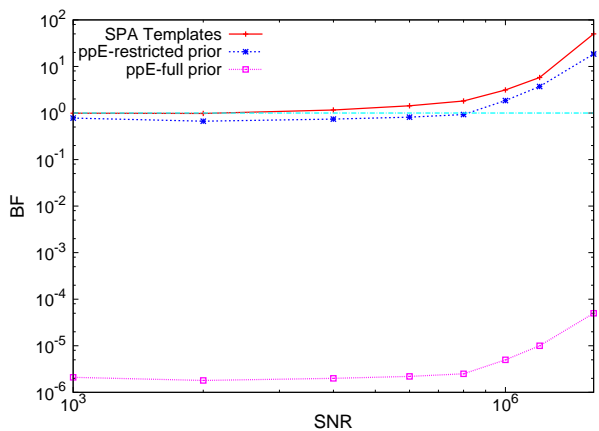} 
\caption{\label{fig:SPASNR} (Color Online) BFs in favor of modified gravity as a function of SNR, calculated by injecting a signal with $\beta_\ST = -3.5$. A BF over 1 indicates a preference for a non-GR theory of gravity. The red (solid) line shows the BFs calculated using SPA templates, and the blue (dashed) line shows those calculated from ppE templates.}
\end{figure}

Figure~\ref{fig:SPASNR} shows the BF as a function of the SNR between GR and either the SPA or the simple ppE templates, using both prior ranges for the ppE templates. Notice that the SPA templates detect the modifications at a much lower SNR than the ppE templates using the full prior, and at an SNR approximately half the value necessary for the ppE templates with the restricted prior. Low SNR, however, is a relative term. The SPA templates detect the GR modifications at SNRs $\approx 6 \times 10^{5}$, which corresponds to a ridiculous luminosity distance of $\approx 10^{3}$ pc (essentially inside the Milky Way). 

Although we expected that the custom-made SPA templates would be more effective than the generic, ppE templates at extracting signals (at sufficiently high SNR), 
it is still worth studying the reason behind this expected result. To do so, we examine the prior and posterior distributions of the two non-GR parameters, $\beta_\ST$ and $\beta_\ppE$, and the \emph{Occam penalty} that arises from each parameter. The Occam penalty is a built-in feature of Bayesian analysis, which causes simple models to be favored over more complicated ones. That is, models with fewer parameters are preferred to models with extra parameters, all else being fixed\footnote{Consider two nested models: $\mathcal{M}_1$ which is parameterized by a single parameter, $\theta$, and $\mathcal{M}_0$ which is unparameterized, i.e.~has $\theta=\theta_0$ where $\theta_0$ is a constant. If the likelihood function for $\mathcal{M}_1$ is a Gaussian, then ${\rm{BF}}_{1,0} \propto (\delta \theta)/\Delta \theta$, where $\delta \theta$ is the characteristic width of the 
posterior in $\theta$, and $\Delta \theta$ is the prior range of $\theta$~\cite{TysonThesis}. Thus, if the value of $\theta$ is entirely unconstrained by the data, there is no penalty for an extra parameter. On the other hand, if $\theta$ is very tightly constrained, there is a large penalty.}.
\begin{figure}[ht]
\includegraphics[clip=true,width= 0.48\textwidth]{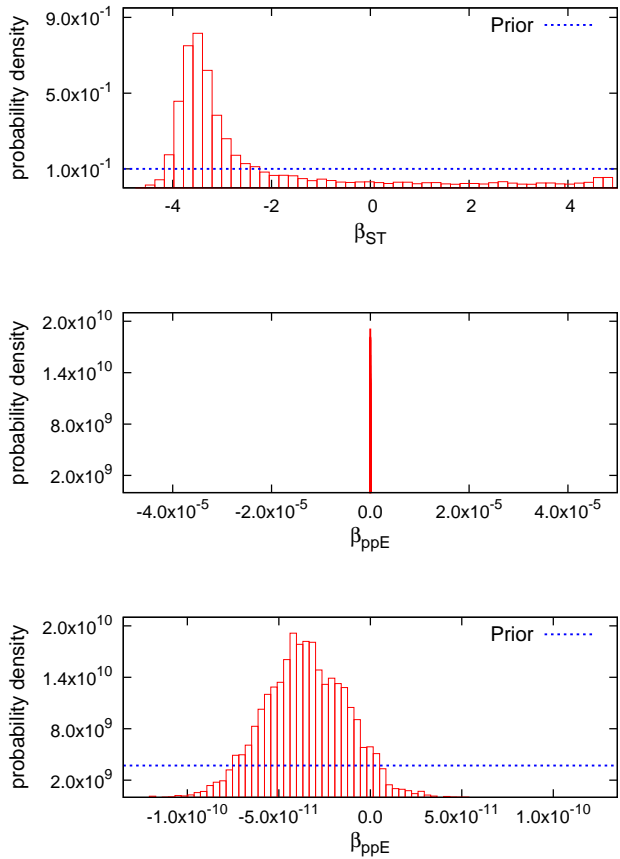} 
\caption{\label{fig:highSNRpost} (Color Online) Posterior distributions for $\beta_\ST$ (top panel) and $\beta_\ppE$ with the full prior (middle panel) and restricted prior (bottom panel), generated by recovering an SNR $600,000$ signal with $\beta_\ppE = -3.5$. Also plotted are the prior densities, although this is not visible in the middle panel. The posterior for $\beta_\ppE$ is so highly constrained compared to the full prior that it is nearly impossible to use this model for the detection of a ST modification to gravity. }
\end{figure}

This brings us to an explanation of the results in Fig.~\ref{fig:SPASNR}. Figure~\ref{fig:highSNRpost} shows the posterior distributions for $\beta_\ST$ and $\beta_\ppE$, using both the full and the restricted priors, for a signal with SNR of $6 \times 10^{5}$, plotted over the entire prior range. The prior distribution is plotted in all three cases, although it is only visible for $\beta_\ST$ and for $\beta_\ppE$ with the restricted prior range. Notice that, where the posterior for $\beta_\ST$ and $\beta_\ppE$ in the restricted case have some weight over most of their entire prior ranges, $\beta_\ppE$ for the full prior range is hugely constrained - so constrained that its posterior distribution looks like a delta function. This means that there is a very large Occam penalty disfavoring this model, and it will take a signal of extremely high SNR to overcome this penalty. As seen from the results in this section, a tighter prior range leads to a smaller Occam penalty, and thus a larger BF in favor of the GR deviation. 

\subsection{Dynamically Scalarized Signals}
\label{dyn-sca-subsec}

Section~\ref{sec:useful-effective-ST} hinted that dynamical scalarization is much more difficult to detect than spontaneous scalarization. In fact, Fig.~\ref{fig:BFvbetaST} shows that the number of effective cycles accrued in dynamically scalarized signals is rather low in general. An example of this is the dynamically scalarized, $\beta_{\ST} = -4.25$ case for a binary with masses $(1.4074,1.7415) M_{\odot}$. Indeed, a Bayesian analysis of this signal shows that such a ST effect is not detectable with simple ppE or 2-parameter ppE templates. 

One way to understand this is by considering the amount of SNR that is accrued in the signal after scalarization has set in. Table~\ref{table:SNR} lists the percentage of the total $\rm{SNR}^2$ contained in the signal \emph{before} at least one of the NSs has become scalarized, for signals with $\beta_\ST = -4.25$. The frequency at which either NS becomes scalarized can be easily extracted by looking at the behavior of the scalar charges as a function of orbital frequency (see Fig.~\ref{fig:fstarvbeta}). From this table, it is clear that there is very little SNR accumulated over the portion of the signal in which scalarization effects are important. Moreover, this SNR accumulation does not occur in the 
frequency region in which the instrument is most sensitive.

\begin{table}[ht]
\centering  
\begin{tabular}{c c c } 
\hline\hline                        
Mass & \quad  Freq.~[Hz] \quad & \quad  $\% \rm{SNR}^2$  \\ [0.5ex] 
\hline                  
$(1.4074,1.7415) M_\odot$ & 314 & 98.1   \\ 
$(1.5145,1.7415) M_\odot$ & 301 & 94.25   \\
$(1.6441,1.7415) M_\odot$ & 286 & 89.6  \\ [1ex]      
\hline 
\end{tabular}
\caption{\label{table:SNR} The percentage SNR squared accrued before dynamical scalarization has begun, for signals with $\beta_\ST = -4.25$ and SNR $\approx 15$. The first column gives the mass of the system, the second the approximate frequency at which dynamical scalarization begins, and the third the percentage of the $\rm{SNR}^2$ accumulated prior to scalarization. Note that for all cases most of the $\rm{SNR}^2$ of the signal is amassed before scalarization becomes significant. }
\end{table}

The reason why the $\beta_{\ST} = -4.25$, dynamically scalarized case cannot be easily detected is precisely that the frequency at which ST modifications become noticeable is rather large. But this frequency is, of course, a function of the masses of the binary and the EoS used. Recall that dynamical scalarization sets in when the energy of the system, roughly speaking the linear combination of the NS compactnesses and the (absolute value of the) gravitational binding energy, exceeds a certain threshold. Therefore, one can imagine a NS binary whose masses and radii are such that the NS compactnesses are very close to exceeding the energy threshold, and thus, dynamical scalarization can set in at very low frequencies, as shown in Fig.~\ref{fig:fstarvbeta}. 

From the effective cycle study illustrated in Fig.~\ref{fig:BFvbetaST}, it appears that the dynamically scalarized binaries that are most easy to detect are those described at the beginning of Sec.~\ref{sec:BF}: a $(1.6441,1.6441) M_{\odot}$ binary at $\beta_{\ST} = -4.5$, a $(1.7415,1.7415) M_{\odot}$ binary at $\beta_{\ST} = -4.25$, and a $(1.5145,1.6441) M_{\odot}$ binary at $\beta_{\ST} = -4.5$. Let us first consider a signal described by the first of these sets of parameters and extract it with a simple ppE template. Such a signal dynamically scalarizes at the lowest frequency of all systems considered (at roughly 80 Hz). This system is indeed detectable, leading to a very large BF and a $\beta_{\ppE}$ posterior that is similar to that shown in the left panel of Fig.~\ref{fig:beta45} for a spontaneously scalarized, $\beta_{\ST} = -4.5$ binary with masses $(1.4074,1.7415) M_\odot$. The width of the $\beta_{\ppE}$  posterior, however, is roughly one order of magnitude larger than in the spontaneously scalarized case, with a variance of $\sigma_{1.6,1.6} = 3.7 \times 10^{-5}$ for the former and $\sigma_{1.4,1.7} = 5 \times 10^{-6}$ for the latter, indicating that the BF in this case is smaller, as expected. 

We can now repeat this analysis for the other two binaries that we expect may be detectable given Fig~\ref{fig:BFvbetaST}. For both cases, we find that $BF \approx 3$, obtained from the Savage-Dickey ratio. This is again in accordance with expectations from Fig~\ref{fig:BFvbetaST}. For the $\beta_\ST = -4.25$ system, the number of effective cycles indicates a marginal detection, which is precisely what we find in this Bayesian analysis. For the $\beta_\ST = -4.5$ system, the number of effective cycles plotted in Fig~\ref{fig:BFvbetaST} suggests the possibility of detection; however, recall that the numbers shown in this figure are upper limits. In this case, our Bayesian analysis indicates that this upper limit is higher (by a factor of $\approx 2$) than the actual number of effective cycles induced by dynamical scalarization of the binary components.

As already mentioned, our analysis thus far has not focused on one important feature of dynamically scalarized signals: the early plunge of the NS binary. That is, once the GW frequency has exceeded the threshold for dynamical scalarization to set in, the NS binary inspirals for a few more cycles, but then plunges and merges soon after. This occurs much earlier than in GR (see e.g.~Fig.~$10$ and~$15$ in~\cite{Palenzuela:2013hsa}). Of course, after the NSs have merged, either a hypermassive NS forms, with a rotating bar that emits GWs at kHz frequencies, or a BH forms, thus cutting out GW emission exponentially through ringdown. The precise form of the waveform during this merger and ringdown phase will depend strongly on the NS equation of state.

We study in an \emph{approximate} fashion whether an early plunge can be detected in a generic modified gravity theory by considering a set of Heaviside signal injections, i.e.~GR  waveforms for which the Fourier amplitude is multiplied by a Heaviside function with 
argument $f^{*}_{\inj} - f$, as we vary the injection cutoff frequency $f^{*}_{\inj} \in (40,10^{3})$Hz. In dynamically scalarized systems, however, the transition from inspiral to early plunge and then merger is \emph{smooth}, while Heaviside templates are clearly not. Therefore, it is obvious that the latter are inappropriate templates to extract realistic dynamical scalarization signals. However, they are good and simple toy-models to study whether an early plunge (and thus an early termination) of the signal could be detected as a non-GR effect in data analysis. Since a smooth transition will be less noticeable than a sharp Heaviside transition, the use of Heaviside templates could be thought of as conservative, i.e.~if a GR deviation cannot be observed with such an abrupt termination, it certainly will not be detectable if the transition is smooth.

We carry out such a study in the following way. We place all such systems at $D_{L} = 
30.75$ Mpc such that the recovered SNR is approximately $15$ when $f^*_{\inj}=100 $ Hz 
and $30$ when $f^{*}_{\inj}=1000$ Hz. For that value of $f^{*}_{\inj}$, the Heaviside signal 
is thus similar to those we have been analyzing throughout this paper and also 
those studied in Ref.~\cite{Sampson:2013lpa}. We then extract such injections with 
templates that exactly match the signal, but with $f^{*}$ included as a template parameter to search over, 
as well as with simple ppE templates with $b=-4$. This value of the exponent parameter is chosen because of its strong correlation with the total mass, which is the parameter that determines the cutoff frequency in GR. As explored in Ref~\cite{Sampson:2013jpa}, the specific value of $b$ that is chosen has little impact on the analysis.

One may worry that approximating the early plunge in this abrupt way may mask the detectability of non-GR effects, as the cycles that are effectively thrown out by this sort of study would contain these effects. Because there is so little SNR contained in those final few cycles of inspiral, however, this should not be an issue. For the systems studied in this paper, there are thousands of orbital cycles before scalarization is activated, and only tens of orbital cycles afterwards. Additionally, for all but a few cases, the orbital cycles that are affected by scalarization occur at a frequency in which the detectors are not very sensitive. These two effects combined mean that there is very little information being discarded by abruptly terminating the waveforms once scalarization has occurred.

Clearly, the earlier the binary plunges (or, in our case, the lower the injection cutoff frequency, $f^*_{\inj}$), the fewer GW cycles the signal will contain in the sensitivity band of the detector. This then translates to a smaller recovered SNR. Thus, in order to detect such a signal at all, we must either be fortunate enough to detect systems that are sufficiently nearby, or fortunate enough to detect enough events such that their stacked SNR is large. Table~\ref{table:SNR-new} shows the luminosity distance required such that the SNR recovered equals 8 for different termination frequencies $f^{*}_{\inj}$. This table also shows what the SNR would have been at such luminosity distances, if the signal did not terminate at $f^{*}_{\inj}$, but rather continued to $1000$Hz. 
\begin{table}[ht]
\centering  
\begin{tabular}{c| c c c c c} 
\hline\hline                        
 $f^{*}_{\inj}$ [Hz] & $100$ &  $65$ &  $51$  &  $44$  &  $39$ \\ [0.5ex] 
\hline                  
$D_{L}$ [Mpc] & $60$ & $30$ & $18$ &  $12$ & $9$  \\ [0.5ex]      
\hline                  
${\rm{SNR}}$ & $15$ & $30$ & $50$ &  $75$ & $100$  \\ [0.5ex]      
\hline \hline 
\end{tabular}
\caption{\label{table:SNR-new} Luminosity distance to the source such that the SNR recovered (up to a threshold frequency $f^{*}_{\inj}$) is equal to 8. The third row shows the total SNR that would be recovered at the given luminosity distances if $f^{*}_{\inj} = 1000$ Hz. Note how rapidly the distance to the source has to be decreased as the threshold frequency is decreased.}
\end{table}

\begin{figure*}[ht]
\includegraphics[clip=true,width=0.45\textwidth]{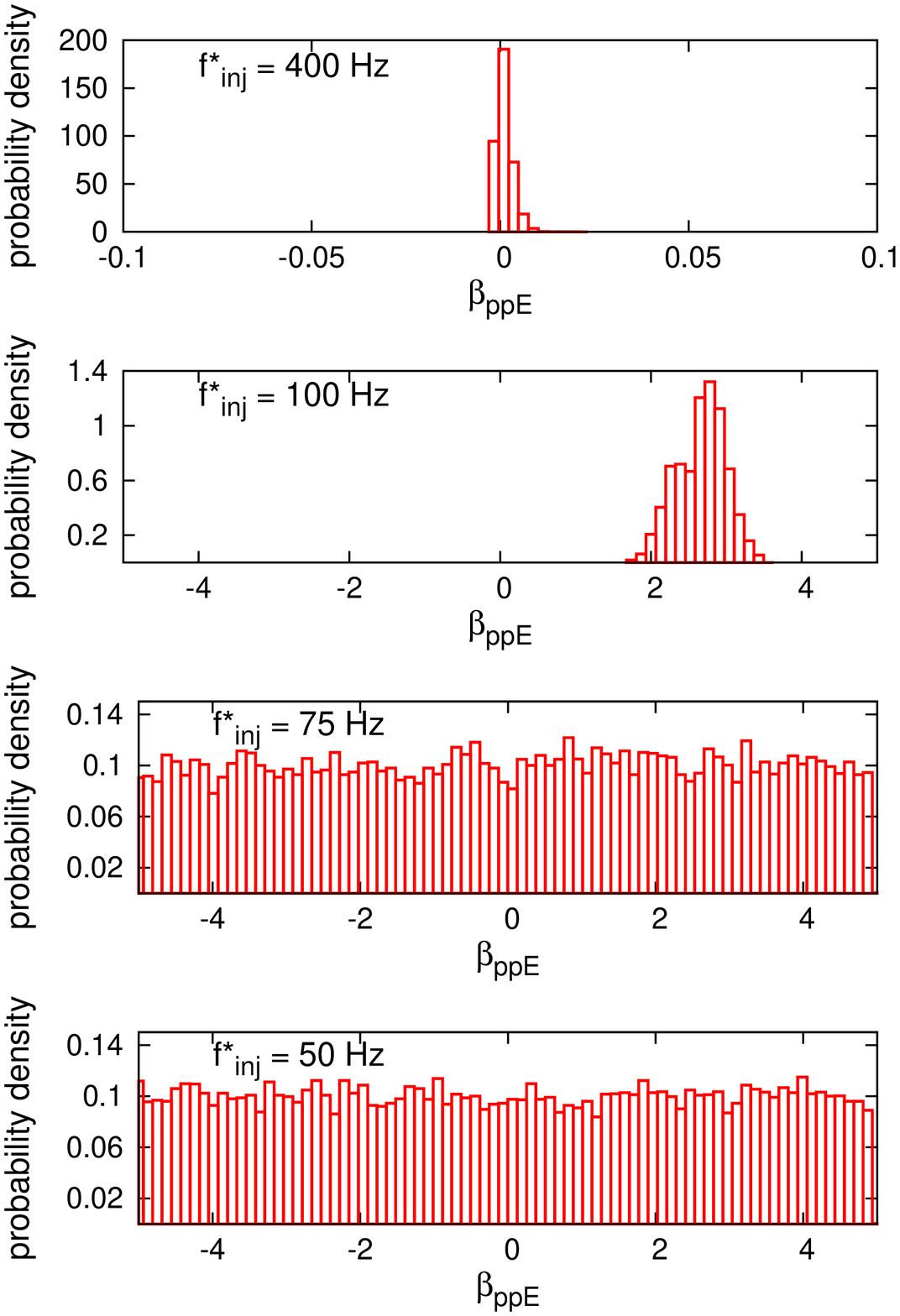} 
\includegraphics[clip=true,width=0.45\textwidth]{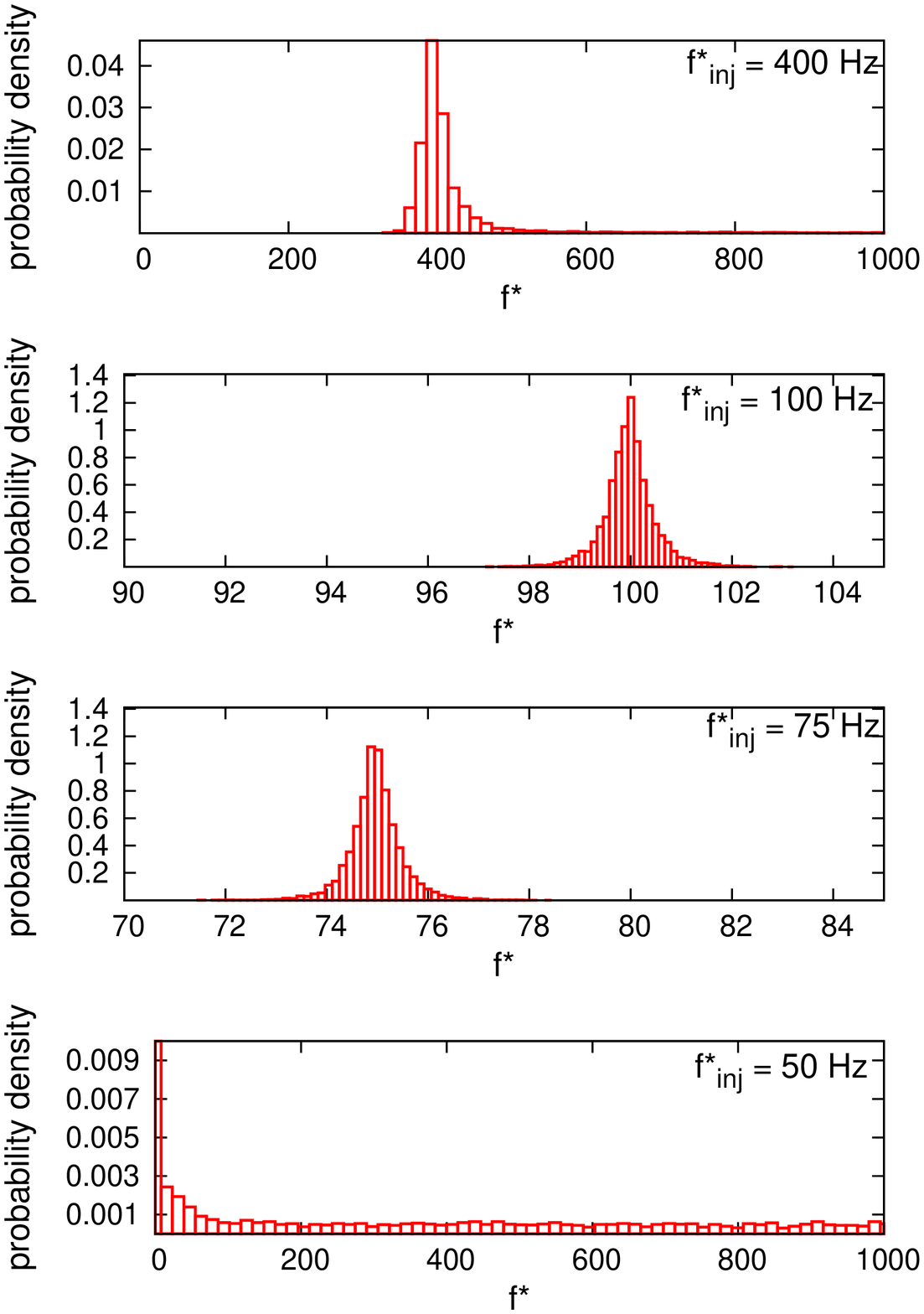} 
\caption{\label{fig:ppEthetaGRtheta} (Color Online) Left panels: posterior distributions for $\beta_\ppE$, recovered by extracting a Heaviside template injection with injected cut-off frequency $f^*_{\inj}$, using simple ppE templates. The only case that shows a strong preference for the non-GR model is for $f^*_{\inj} = 100$ Hz. Right panels: posterior distributions for the recovered $f^*$, generated by extracting a Heaviside injection with templates of the same family and different injected cut-off frequencies $f^{*}_{\inj}$. Here, the cases with $f^*_{\inj}=100$ Hz and $f^*_{\inj}=75$ Hz are distinguishable from GR.}
\end{figure*}

Figure~\ref{fig:ppEthetaGRtheta} shows the posterior distributions for the recovered values of $f^*$ using Heaviside templates, and the posteriors for the recovered values of $\beta_\ppE$ using simple ppE templates. All injections focus on a $(1.6,1.6) M_\odot$ NS binary, with the same polarization angle and sky position as all other injections in this paper, and $D_L = 30.75$ Mpc. The prior range on the search parameter $f^*$ is uniform between $0$ to $1000$ Hz for the Heaviside templates, and thus, the prior density is $1/1000 = 0.001$. Recall that the prior range on the search parameter $\beta_{\ppE}$ is also uniform with range $-5$ to $5$ for the simple ppE templates, and thus, the prior density is $1/10 = 0.1$ in this case. 

The interpretation of the posterior distributions for the recovered values of $f^*$ as tests of GR is somewhat subtle, because it is not entirely clear what the ``GR value'' for $f^*$ should be. In principle, the inspiral should end when the NSs begin to plunge, and certainly by the time the stars have come into contact. The GW frequency of the latter, $f_{\rm cont}$, depends both on the component masses and the EoS; for a $(1.6,1.6) M_\odot$ binary it is between $1250$ and $2050$ Hz, depending on the NS radius.  As a simple and practical measure of the plunge, one could choose the GW frequency at the innermost stable circular orbit (ISCO) of a test particle in a Schwarzschild spacetime: $f_\ISCO = 6^{-3/2}/(\pi M)$; for a $(1.6,1.6) M_\odot$ NS binary, $f_\ISCO = 1354$ Hz. This is a suitable measure for the beginning frequency of plunge in GR, provided the NS is compact enough such that its contact frequency is above $f_{\ISCO}$. For the systems we study here,  $f_{\rm cont}>f_{\ISCO}>1000$ Hz, where recall that the latter is the highest frequency of integration in all cross-correlations. We can therefore take the GR value of $f^{*}$ to be $1000$ Hz, and calculate the BF by comparing posterior and prior densities at this point.

The right panels of Fig.~\ref{fig:ppEthetaGRtheta} show that the Heaviside templates are able to distinguish a deviation from GR provided the injected cutoff frequency $f^{*}_{\inj}$ is above $\approx 50$ Hz and below $\approx 400$ Hz. These panels present the posterior distributions for the parameter $f^*$, recovered using Heaviside templates on injections with various values of $f^*_{\inj}$. The ability of Heaviside templates to distinguish GR deviations can be established by computing the BFs for each of these panels through the Savage-Dickey density ratio (recall that the BF is the ratio of the posterior to the prior density at the GR value of the search parameter, $f^*=1000$ Hz in the Heaviside template case). For instance, the ${\rm{BF}} \approx 1$ when $f^*_\inj = 50$ Hz, while the ${\rm{BF}} \approx 5$ when $f^*_{\inj} = 400$ Hz, a marginal detection of a GR deviation.

The left panels of Fig.~\ref{fig:ppEthetaGRtheta} show that the ppE templates can also distinguish Heaviside-type deviations from GR, but this time provided $f^{*}_{\inj}$ is above $\approx 75$ Hz and below $\approx 400$ Hz. These panels present the posterior distributions for $\beta_\ppE$, recovered using simple ppE template with $b=-4$ on Heaviside injections with various values of $f^{*}_{\inj}$. The BF can still be computed through the Savage-Dickey density ratio, except that now GR is recovered when the value of $\beta_{\ppE}$ is zero, and recall that the prior density is $0.1$. For instance, the ${\rm{BF}} \approx 1$ when $f^*_\inj = 75$ Hz, while the BF is clearly much larger than unity when $f^{*}_{\inj} = 100$ Hz. 

Why is detectability of a GR deviation difficult for very low or very large $f^{*}_{\inj}$? For very low injected cutoff frequencies (e.g.~below $50$Hz for the Heaviside templates and $75$ Hz for the ppE templates), the analysis fails to detect a signal altogether -- a reasonable result, considering that these very low injected cutoff frequencies drop the total recovered SNR of the signal below 8. For very high injected cutoff frequencies (e.g.~above $400$ Hz), the deviation from GR occurs too far outside of the detector's most sensitive band to be noticeable. Put another way, not enough SNR is accrued while the GR deviation is active. 

Although the choice of a signal at $D_{L} = 30.75$ Mpc leads to a recovered SNR $\approx 30$ when $f^*_{\inj}=1000$ Hz, a reasonable choice for comparison to other results in this paper, systems at such a close distance are not very likely. A more reasonable expectation is a system at twice that luminosity distance, $D_L \approx 61.5$ Mpc, such that the signal that \emph{would} have total SNR of $\sim15$ if $f^{*}_{\inj}=1000$ Hz, but which, if subject to the early plunges analyzed here, is in fact a signal with lower recovered SNR. When we inject Heaviside signals at such a $D_{L}$, so that the SNR $\sim 15$ if $f^{*}_{\inj} = 1000$ Hz, we find results qualitatively similar to those described in the previous paragraph, but with a narrower detectable injected cutoff frequency range. The high injected cut-off frequency for detectability drops to $\approx 300$ Hz, and the low cut-off frequency rises to $\approx 90$ Hz. This result is in accordance with expectations.

The ability to detect the early plunge of a binary system may have an important implication in terms of the detectability of dynamical scalarization. As we have seen in Fig.~\ref{fig:fstarvbeta}, higher mass NS binaries can dynamically scalarize at frequencies below $400$Hz for $\beta_{\ST} \approx -4$. This then suggests that the inclusion of a early plunge, the merger and the post-merger phase in a data analysis study may allow for constraints on ST theories at such values of $\beta_{\ST}$. Notice, in particular, that such constraints are stronger than those obtained when including only the inspiral phase. 

One must be very careful, however, when extrapolating results and promptly concluding that the inclusion of the plunge and merger portions of these signals will increase their distinguishability. First, there are very few cycles in the post-plunge phase, and thus very little SNR accumulated after the plunge. Second, and perhaps more importantly, the post-plunge phase will be strongly affected by the NS equation of state. Degeneracies between equation-of-state effects and non-GR effects imply that the post-plunge phase of NSs may not be as useful as a means to test GR. Of course, a more detailed analysis is required to derive solid conclusions.

\section{Conclusions}
\label{sec:conclusions}
In this paper, we sought to answer one overarching question: can deviations from GR in GW signals that are caused by a certain class of ST theories be detected with aLIGO-type instruments? We find that this is the case, irrespective of whether the deviation arises from spontaneously/induced or dynamically scalarized NSs. These projected constraints will be complementary and at least comparable to current binary pulsar ones. 

Not all spontaneous/induced and dynamical scalarization effects, however, are easily detectable. For a spontaneous/induced scalarization to be detectable, the scalar field anchored on each NS must be large enough, and the masses sufficiently dissimilar, that dipole radiation becomes important. This occurs, for example, for values of $\beta_{\ST} \approx -4.5$ and a NS binary with masses $(1.4,1.7) M_{\odot}$, for the simple polytropic EoS that we consider in this paper (with a different EoS potentially changing these values). For a dynamically scalarized effect to be detectable in aLIGO, such scalarization must occur at a sufficiently small GW frequency so that enough GW cycles are affected in a frequency region that detectors are sufficiently sensitive to. This occurs, for example, for values of $\beta_{\ST} \approx -4.5$ and a NS binary with masses $(1.6,1.6) M_{\odot}$, which scalarizes at a GW frequency of roughly 80 Hz
(again with some variability of these values depending on the EoS). 

We additionally investigated the detectability of another effect associated with dynamical scalarization: the early plunge that would be induced in binaries that undergo such scalarization. We found that the sudden cutoff of a GW signal at frequencies below those expected in signals described by GR is detectable using both simple ppE templates and GR templates with a Heaviside function, for certain ranges of cutoff frequencies. Such a plunge, for example, could occur for values of $\beta_{\ST} = -4$ and a NS binary with masses $(1.7,1.7) M_{\odot}$. Of course, an early plunge cannot be accurately modeled through a Heaviside function, and so a more careful data analysis study that includes inspiral-merger-post-merger signals is needed.  

In the process of reaching these answers, we explored a 
measure that can help estimate whether different types of phase effects are 
detectable with GWs: the effective cycles of phase. We found that for a wide 
variety of different non-GR signals, the modification to the GR phase needs to 
lead to roughly 4 cycles of effective phase to be detectable with an aLIGO 
detector at SNR 15. We further found that such detectability is independent of whether one 
uses custom-made ST templates or model-independent templates to search for 
deviations at the expected SNRs of aLIGO-type detectors. 

A final question that we considered was whether custom-made, theory-specific templates are more useful at detecting deviations from GR than the model-independent ppE template family. We answered this question by looking at both extremely high SNR signals with ST signatures that were undetectable at low SNR, and by looking at low SNR signals with ST signatures that were easily detectable at reasonable SNR. In both cases, we found that ppE templates perform almost as well as custom-made, SPA templates at distinguishing non-GR signals from GR ones. 

Future work could concentrate on extensions of the analysis presented here. One interesting extension would be to repeat this work for NSs with realistic EoSs. The work in Ref.~\cite{Barausse:2012da,Palenzuela:2013hsa}, which we used in this paper exclusively, used a polytropic EoS, but this is easily generalizable to more realistic EoSs (c.f.~Ref.~\cite{Shibata:2013pra}). One may find EoSs and masses that do not lead to spontaneous/induced scalarization for binary pulsars, yet lead to spontaneous/induced scalarization for systems that can be detected with GWs. Such systems would thus evade binary pulsar constraints and yet potentially lead to detectable deviations with aLIGO. Again, it is important to emphasize that stiffer/softer EoSs will lead to qualitatively (and quantitatively) different behavior. A very stiff EoS could perhaps support more compact stars that spontaneously scalarize at lower frequencies. A systematic study of these effects is left for future consideration.

Another interesting analysis would be to study whether one can find EoSs or masses for which dynamical scalarization sets in at very low frequency, e.g.~close to 10Hz. Reference~\cite{Sampson:2013jpa} estimated that for an abrupt GR modification to be detectable with an aLIGO-like detector at SNR 10, such a modification would have to start at a GW frequency below 100 Hz. Most cases of dynamical scalarization that we explored occur at $\sim$200Hz or higher, but there are some instances in which scalarization occurs at a lower frequency, and it is possible that different mass/EoS combinations would produce more of these scenarios.

Other extensions may include adding more complexity to the signals and the templates, through the inclusion of the merger and post-merger phases,  as well as the inclusion of spin~\cite{Klein:2013qda,Chatziioannou:2013dza,Chatziioannou:2014bma,Chatziioannou:2014coa} and eccentricity~\cite{Yunes:2009yz} effects. 
Recall that with respect to second-generation, ground-based detectors it is common to regard the 
NS merger and post-merger phases as unimportant for testing GR as they occur at kHz frequencies where such detectors
are  least sensitive.
Dynamically scalarized NSs, however, could plunge at much lower frequencies, and such effects may be detectable. References~\cite{Chatziioannou:2014bma,Chatziioannou:2014coa} showed that the inclusion of spin in NS binaries can have a large effect on parameter estimation. Similar conclusions were arrived at when including more complexity in GW signals to test GR (see e.g.~\cite{Stavridis:2010zz,Yagi:2009zm,Yagi:2009zz}). Also, Ref.~\cite{Palenzuela:2013hsa} showed that eccentric NS binaries in ST theories can give rise to scalarization/descalarization phenomena that may affect the binary's orbital evolution (effectively decreasing the eccentricity faster than in GR) at sufficiently low frequencies to be detected. 

Another interesting avenue for future work would be to repeat the analysis of this paper but with a non-template based search algorithm that may be more sensitive to the post-merger phase. In our analysis, we used a template-based search of inspiral signals, neglecting the post-merger phase that occurs at high GW frequencies where aLIGO's sensitivity will be weaker. Recently, however, Ref.~\cite{Clark:2014wua} used a template-free burst algorithm to show that the merger phase may be sufficiently detectable by aLIGO to discern between a prompt collapse scenario and the formation of a hypermassive NS, if the event occurs at $\approx 10 \; {\rm{Mpc}}$. Dynamical scalarization will modify the post-merger phase, also leading to either prompt collapse or hypermassive NS formation, depending on the masses of the binaries. Such effects, however, will probably be somewhat degenerate with the EoS, and thus, it is unclear whether merger modifications will be strong enough to detect a GR deviation. 

One final way to study the robustness of our conclusions would be to consider GW detection with a network of GW detectors, with second-generation detectors and noise tuning, with a large number of second-generation detections and stacking~\cite{Berti:2011jz,DelPozzo:2013ala}, or with future GW detectors, such as the Einstein Telescope~\cite{et}. All of these could lead to much lower noise at GW frequencies around 100 Hz, which would then have multiple effects. First, better sensitivity at 100Hz should push the threshold frequency at which GR deviations can be detected to higher frequencies. Second, better sensitivity overall should lead to individual detections with higher SNR and to a higher number of detections per year. Combining all of this, and perhaps through the use of stacking, one may be able to detect dynamical scalarization for lower values of $|\beta_{\ST}|$. One should keep in mind, however, that the direct detection of the additional breathing mode will be extremely hard, even when detecting GWs with multiple instruments, since in ST theories that pass Solar System tests, the interaction of such modes with a detector is suppressed by $\psi_0$~\cite{Barausse:2012da}, which is constrained to $\lesssim10^{-2}$ because of the Cassini bound. 

Finally, let us address the relationship between our work and that reported in Ref.~\cite{Taniguchi:2014fqa}, which appeared after the submission of this paper. One may be led to believe that the conclusion arrive at in this paper and those of Ref.~\cite{Taniguchi:2014fqa} are not in agreement, with respect to the detectability of scalarization effects with aLIGO. Reference~\cite{Taniguchi:2014fqa} first calculates the total number of cycles of phase that are accumulated in a GW signal due to the presence of scalarization effects. They then note that this number is larger than or comparable to the total number of cycles of phase that will be accumulated due to EoS effects within GR. Since the latter may be detectable with next generation detectors~\cite{Damour:2012yf,Wade:2014vqa,DelPozzo:2013ala}, they then argue that scalarization effects may also be detectable. This conclusion is in fact in \emph{agreement} with our findings because EoS effects can be measured only provided the SNR is sufficiently high (roughly above 30). Our analysis used SNRs in the tens, as expected from the first few years of detection; for such signals, non-GR effects are not detectable. Reference~\cite{Taniguchi:2014fqa} also finds that scalarization can be detected when it occurs at GW frequencies larger than roughly 130 Hz, which is in perfect agreement with our findings (once one converts orbital to GW frequency). 

Although Ref.~\cite{Taniguchi:2014fqa} does not claim that the total cycles of phase due to a particular effect can directly tell us about detectability, it is possible to misread the conclusions of this paper to indicate that they can. We therefore emphasize again here that the total cycles of phase have no analytic connection to the Bayes factor and, in fact, fail to account for parameter covariances. This affects detectability in two main ways. One is obvious - a non-GR phase term that has large correlations with other system parameters will be easy to fit using GR templates. This will make the effect very difficult to discern, as illustrated in Fig.~\ref{fig:BFNuse}. The other, less obvious consequence is that although effects that enter at high PN order accumulate more slowly, they may be detectable when they lead to smaller numbers of phase cycles than those that enter at low PN order. This is again directly due to parameter covariances. In the case of spontaneous or dynamical scalarization, non-GR effects that enter at low PN order will have high covariances with system parameters, such as the chirp mass. Higher PN order effects will have weaker covariances, but they lead to much weaker effects, with most of the total dephasing accumulating from the low PN order terms that have large covariances. Thus, the use of total cycles of phase to claim detectability of non-GR effects is not appropriate and one should really either use the effective cycles discussed in this paper or a full Bayesian analysis.

Also, Ref.~\cite{Taniguchi:2014fqa} (at least in its first arXiv version) states that the simulations of Ref.~\cite{Barausse:2012da} ``misread'' the output of the LORENE code used to determine the simulation initial data, and claims that the gravitational masses given in Ref.~\cite{Barausse:2012da} are incorrect. This is certainly not the case and the gravitational masses we give in Ref.~\cite{Barausse:2012da} are exactly those of the LORENE output. As it is well known, there is no unambiguous way of defining individual gravitational masses for a tight binary system within General Relativity. We provided the readers of Ref.~\cite{Barausse:2012da} with the exact LORENE-given gravitational masses in order to clearly identify our initial data and ensure our results will be reproducible by others. An
alternative, which is the one followed in Ref.~\cite{Taniguchi:2014fqa}, would have been to give the gravitational masses of the stars in isolation, but this may have caused unnecessary confusion, as the gravitation masses are never used in Ref.~\cite{Barausse:2012da} except for identifying the initial data. 

Reference~\cite{Taniguchi:2014fqa} also raises question on the dynamics of the most massive case presented in Ref.~\cite{Barausse:2012da} and used here. Their concern is related to our choice of initial data for such already scalarized case, and is about whether this could cause an earlier plunge. Such concern was already addressed in Ref.~\cite{Palenzuela:2013hsa}, where we \textit{(i)} show the results of simulations with large initial separations (e.g. Figs.~4 and 5, where no plunge is present at large separations) and \textit{(ii)} validate our simulations with an enhanced PN model. 
Moreover, we stress here that the initial data used in Ref.~\cite{Barausse:2012da} for the low-mass case with dynamical scalarization are exact (because in the absence of spontaneous scalarization, the ST initial data are the same as in GR).

\acknowledgments
We thank Gilles Esposito-Farese and Gabriela Gonzalez for 
insightful conversations.
NY acknowledges support from NSF grant PHY-1114374 and the NSF CAREER
Award PHY-1250636, as well as support provided by the National
Aeronautics and Space Administration from grant NNX11AI49G, under
sub-award 00001944. LS and NJC acknowledge support from NSF grant PHY-1306702.
EB acknowledges support from
the European Union's Seventh Framework Programme (FP7/PEOPLE-2011-CIG)
through the  Marie Curie Career Integration Grant GALFORMBHS PCIG11-GA-2012-321608.
AK is supported by NSF CAREER Grant No. PHY-1055103.
LL acknowledges support by NSERC through a Discovery Grant and CIFAR. LL thanks 
the Institut d'Astrophysique de Paris and the ILP LABEX (ANR-10-LABX-63),
for hospitality during a visit supported through the Investissements d'Avenir programme under
reference ANR-11-IDEX-0004-02.
Research at Perimeter Institute is supported through Industry Canada
and by the Province of Ontario through the Ministry of Research and Innovation.

\appendix
\section{Scalar-Tensor Theories with $\beta_\ST <0$ \label{AppA}}

In this appendix, we discuss how ST theories with $\beta_{\ST} < 0$ repel cosmological solutions away from their GR counterparts. 
This will be done by essentially repeating the analysis in Refs.~\cite{PhysRevLett.70.2217,PhysRevD.48.3436} (see also Ref.~\cite{cosmoST}), but flipping the sign of $\beta_{\ST}$. 

First, we review the work of Refs.~\cite{PhysRevLett.70.2217,PhysRevD.48.3436}. Assuming a ST theory of the type discussed in this paper, 
the Friedmann-Robertson-Walker evolution of the scalar field is described by~\cite{PhysRevLett.70.2217,PhysRevD.48.3436} 
\be
\frac{2}{3 - \varphi '^2}\varphi'' + (1-w)\varphi ' = -(1-3 w) \bar{\alpha}(\varphi)\,.
\label{Eq:diffeq}
\ee
Here, $ \varphi\equiv \psi \sqrt{4 \pi G}$, primes indicate dimensionless derivatives with respect to a time variable $\tau$ such that 
$d\tau = H_{\rm E}(t_{\rm E}) dt_{\rm E}$, or simply $\tau=\ln a_{\rm E}(t_{\rm E})+$ const, where $a_{\rm E}$, $H_{\rm E}$ and $t_{\rm E}$ 
are the expansion parameter, the Hubble expansion rate and the cosmological time in the Einstein frame respectively. Also, 
$\bar{\alpha}(\varphi) = \partial \ln(1/\phi(\varphi))/\partial \varphi \equiv \partial a(\varphi)/\partial \varphi$,  
but in the theories of interest to us, $\bar{\alpha}(\varphi) = \beta_{\ST} \varphi$. Finally, in this equation, $w=p/\rho$ 
is the usual cosmological EoS parameter, where $p$ and $\rho$ are the pressure and density of the cosmic fluid respectively, 
i.e.~$w = -1$, $1/3$, or $0$ during the inflationary, radiation and matter eras respectively. Clearly, Eq.~\eqref{Eq:diffeq} 
can intuitively be understood as a particle with (velocity-dependent) mass $m = 2/(3- \varphi'^2)$ moving in a potential, $a(\varphi)$, with a ``friction'' term $(1-w)\varphi'$.

During the inflationary era ($w=-1$), Eq.~\eqref{Eq:diffeq} admits a solution $\varphi={\varphi}_{0}+\sqrt{3} \tau$ , as can be checked explicitly\footnote{In fact, this is a solution for any $w$.} by replacing it into Eq.~\eqref{Eq:diffeq} multiplied by $3 - \varphi '^2$. We now show that, during inflation, this solution is indeed an attractor at linear order. To this purpose, let us write $\varphi= \sqrt{3} \tau+\delta \varphi$ (where we have absorbed ${\varphi}_{0}$ in $\delta \varphi$). At linear order in $\delta \varphi$, Eq.~\eqref{Eq:diffeq} yields 
\be
\delta \varphi''-6 (1 + 2 \beta_{\ST} \tau) \delta \varphi' =0\,,
\ee
which in turn gives
\begin{multline}
\delta \varphi(\tau)=\varphi_1+\frac{\sqrt{\frac{\pi }{2}} e^{-\frac{3}{2 \beta_{\ST}}} \left(\sqrt{3} \varphi_2-3\right) 
}{6 \sqrt{-\beta_{\ST}}}
\\\times\left[-\rm{Erf}\left({\sqrt{\frac{3}{-2 {\beta_{\ST}}}} (2 \beta_{\ST}
   \tau+1)}\right)+\rm{Erf}\left({\sqrt{\frac{3}{-2 \beta_{\ST}}}}\right)\right]
  \end{multline}
where ${\rm{Erf}}[\cdot]$ is the error function and we have chosen the initial conditions $\varphi_{1}=\varphi(0)$ and $\varphi_{2}=\varphi'(0)$, so as to match respectively the value of $\varphi=\sqrt{3} \tau+\delta \varphi$ and that of its first time derivative at $\tau=0$. Because ${\rm{Erf}}(x)\to 1$ as $x\to \infty$, it is
clear that $\delta \varphi$ goes asymptotically to a constant, hence the asymptotic solution becomes $\varphi= \sqrt{3} \tau+\delta \varphi\approx \varphi_0+ \sqrt{3} \tau$.

During the radiation era, $w=1/3$ and the general solution to Eq.~\eqref{Eq:diffeq} is~\cite{PhysRevLett.70.2217,PhysRevD.48.3436}
\be
\varphi=\varphi_\infty - \sqrt{3} \ln\left[K e^{-\tau} + (1 + K^2 e^{-2 \tau})^{1/2}\right]
\ee
where $\varphi_\infty$ and $K$ are integration constants. This shows that a non-zero initial value of $\varphi$ is damped away in the
radiation era. This can also be seen by solving  Eq.~\eqref{Eq:diffeq} under the approximation $ \varphi'\approx0$, which yields 
$\varphi'' +\varphi ' = 0$ and thus $\varphi \approx \varphi_\infty-K e^{-\tau}$.

Finally, in the matter era $w=0$, the solution $\varphi={\varphi}_{0}+\sqrt{3} \tau$ to Eq.~\eqref{Eq:diffeq} is still an attractor. To prove that this is so, we write $\varphi= \sqrt{3} \tau+\delta \varphi$ (where again we have absorbed ${\varphi}_{0}$ in $\delta \varphi$), which together with Eq.~\eqref{Eq:diffeq} yields 
\be
\delta \varphi''-3 (1 + \beta_{\ST} \tau) \delta \varphi' =0
\ee
at linear order in $\delta \varphi$.
Solving this equation, one obtains
\begin{multline}\label{matterera}
\delta \varphi(\tau)=\varphi_1+\frac{\sqrt{\frac{\pi }{2}} e^{-\frac{3}{2 \beta_{\ST}}} \left(\sqrt{3} \varphi_2-3\right) 
}{3 \sqrt{-\beta_{\ST}}}
\\\times\left[-\rm{Erf}\left({\sqrt{\frac{3}{-2 {\beta_{\ST}}}} (\beta_{\ST}
   \tau+1)}\right)+\rm{Erf}\left({\sqrt{\frac{3}{-2 \beta_{\ST}}}}\right)\right]\,,
  \end{multline}
where $\varphi_{1}=\varphi(0)$ and $\varphi_{2}=\varphi'(0)$ are the two constants of integration, chosen
to match respectively the value of $\varphi=\sqrt{3} \tau+\delta \varphi$ and that of its first time derivative at $\tau=0$. Again, this shows that  $\varphi={\varphi}_{0}+\sqrt{3} \tau$ is
a linear attractor for  Eq.~\eqref{Eq:diffeq}. Because in the radiation era preceding the matter era $\varphi$ is exponentially small,
we can assume $\varphi_{1}=\varphi(0)\approx0$ and $\varphi_{2}=\varphi'(0)\approx0$ (setting $\tau=0$ at the end of the radiation era). Inserting these
conditions into Eq.~\eqref{matterera}, we obtain $\varphi(\tau_{\rm now})=\sqrt{3}\tau_{\rm now}+\delta \varphi(\tau_{\rm now})\approx16$
at the present time $\tau_{\rm now}\approx 10$~\cite{PhysRevLett.70.2217,PhysRevD.48.3436}  for $\beta_{\ST}=-4.5$.

With the present value of the scalar field at hand, we can now study what effect this has on Solar System tests. First, we relate the ppN parameter, $\gamma_{\rm ppN}$, to $\bar{\alpha}$ via~\cite{PhysRevLett.70.2217} 
\be
1-\gamma_{\rm ppN} = \frac{2 {\bar{\alpha}}^2}{1 + \bar{\alpha}^2} = \frac{2 \beta_\ST^2 \varphi^2}{1 + \beta_\ST^2\varphi^2}.
\ee
In GR, $1-\gamma_{\rm ppN} = 0$ and Solar System observations have placed stringent bounds on this quantity: $|1-\gamma_{\rm ppN}| < 2\times10^{-3}$~\cite{PhysRevLett.70.2217}. 
Evaluating $1-\gamma_{\rm ppN}$ at the present time  for $\beta_{\ST}=-4.5$ (i.e.~$\varphi=\varphi_{\rm now}\approx 16$), we obtain $1-\gamma_{\rm ppN} \approx 2$, which is clearly in violation of Solar System experiments. In fact, since $\varphi$ will continue to linearly grow ever larger, $\gamma_{\rm ppN}$ will continue to approach $2$ as  
\be
1-\gamma_{\rm ppN} \approx 2 \left[1 - \frac{1}{3 \beta^{2} \tau^{2}} + {\cal{O}}\left(\frac{1}{\tau^{4}}\right) \right]\,,
\ee
Note, however, that a different functional form for $\bar{\alpha}(\varphi)$ and/or the presence of a suitable potential for the scalar field
may lead to different behavior, 
but such issues have not yet been fully explored (but see Ref.~\cite{cosmoST} for some work in this direction).

\section{Relationship between $\mathcal{N}_u$ and $\mathcal{N}_e$ \label{AppB}}
The useful cycles of phase, $\mathcal{N}_u$,  and the effective cycles of phase, $\mathcal{N}_e$, can be related (in the limit of small dephasings) by a simple calculation. Via this calculation, in this appendix we show that the relationship between these two quantities is dependent only on the PN order of the difference in phase, focusing only on the inspiral phase with the PN approximation. 

Consider two GW signals whose phases differ only by a ppE term of the form $\beta u^b$. That is, $\Phi_1 = \Phi_\GR$ and $\Phi_2 = \Phi_\GR + \beta u^b$. We can then write $\mathcal{N}_u$ as
\be
\mathcal{N}_u = \frac{{\rm SNR}^2}{\int \frac{h_c^2}{Sn(f)} \frac{1}{N(f)}d \ln f} -  \frac{{\rm SNR}^2}{\int \frac{h_c^2}{Sn(f)} \frac{1}{N(f) + \delta}d \ln f},
\ee
where $\delta = \Phi_1 - \Phi_2 = \beta u^b$ in this case. In the limit that $\delta$ is small, we can expand this expression to get

\be
\mathcal{N}_u = -{\rm SNR}^2 \frac{\int \frac{h_c^2}{Sn(f)}\frac{\delta}{N_\GR^2}d\ln f}{\left(\int \frac{h_c^2}{Sn(f)}\frac{1}{N_\GR}d \ln f \right )^2}.
\label{Eq:Nuseexp}
\ee
Given the form of $\delta$, it is clear that $\mathcal{N}_u$ can be written as 
\be
\mathcal{N}_u = \beta g(b),
\ee
where $g(b)$ is the function defined by performing the integrations in Eq.~\eqref{Eq:Nuseexp}.

Similarly, we can write the effective cycles of phase
\be
\mathcal{N}_e = \frac{\left(\int \frac{h_c^2}{Sn(f)} \delta^2 d\ln f \right)^{1/2}}{2\pi {\rm SNR}} = \beta h(b),
\ee
where $h(b)$ encapsulates the b-dependence of the integral in the above equation. 

Taking the ratio of these two expressions, it is also clear that all $\beta$ dependence cancels exactly, and the two are related by some complicated function in $b$, the exponent parameter.

\bibliography{master}
\end{document}